\documentclass[fleqn,usenatbib]{mnras}

\usepackage{newtxtext,newtxmath}
\usepackage{float}
\usepackage{multirow}
\usepackage{amsmath}
\usepackage[T1]{fontenc}
\usepackage{graphicx}	
\usepackage{booktabs}

\DeclareRobustCommand{\VAN}[3]{#2}
\let\VANthebibliography\thebibliography
\def\thebibliography{\DeclareRobustCommand{\VAN}[3]{##3}\VANthebibliography}

\title[CARPP: Radiative Transfer \& Profile Parameters]{
  CARPP: Parametric Radiative-Transfer Fitting of Molecular Cores from Dust Continuum Data
}

\author[Y. Xing et al.]{
  Yuchen Xing,$^{1,2}$
  Di Li,$^{1,3}$\thanks{E-mail: dili@tsinghua.edu.cn}
  Nannan Yue,$^{1}$
  Zhiyuan Ren,$^{1}$
  Qizhou Zhang,$^{4}$
  Sihan Jiao,$^{1,5}$
  \newauthor
  Xin Lyu,$^{1,2}$
  Jiawei Liu$^{1,2}$
\\
$^{1}$National Astronomical Observatories, Chinese Academy of Sciences, Beijing 100101, China\\
$^{2}$University of Chinese Academy of Sciences, Beijing 100049, China\\
$^{3}$Department of Astronomy, Tsinghua University, Beijing 100084, China\\
$^{4}$Center for Astrophysics | Harvard \& Smithsonian, 60 Garden Street, Cambridge, MA 02138, USA\\
$^{5}$Max Planck Institute for Astronomy, Konigstuhl 17, D-69117 Heidelberg, Germany
}

\date{Accepted XXX. Received YYY; in original form ZZZ}

\pubyear{\the\year}

\begin{document}
\label{firstpage}
\pagerange{\pageref{firstpage}--\pageref{lastpage}}
\maketitle

\begin{abstract}
The density profiles of dense molecular cores are important indicators of their physical and evolutionary states. Multi-wavelength dust continuum data offers excellent constraints on the density profile of cores. Here we introduce CARPP (Core Analysis via Radiative Transfer and Profile Parameters), a publicly available fitting package that generates optimized core density and temperature profiles based on parameterized radiative transfer calculations. CARPP assumes spherical symmetry and adopts physically motivated parametric forms for the density and temperature profiles, and uses dust continuum data for fitting. Tests on synthetic data show that CARPP achieves high accuracy, namely averaged relative errors of CARPP's seven parameters being $<20\%$, when the data quality satisfies $\frac{\rm RMS \,\, noise}{[\rm peak \,\, flux]} < 0.025\times \frac{[r_0]}{\rm resolution} +0.05$, where $r_0$ is the core's characteristic radius. We select the low-mass core TMC-1C and the high-mass core Ori2-2 to demonstrate CARPP's performance on real data. It classifies TMC-1C as a Bonnor-Ebert sphere in near-hydrostatic equilibrium, while Ori2-2 exhibits a power-law-dominated profile indicative of a collapsing envelope. This capability establishes CARPP as a powerful and versatile tool to classify the dynamical states of individual cores. It offers an optimal balance between physical fidelity and computational efficiency, serving as a practical, standardized alternative to both over-simplified SED analyses and complex, time-intensive 3D radiative-transfer modeling.
\end{abstract}


\section{Introduction}

Dense molecular cores are the sites of star formation. Their physical structures are closely linked to the efficiency and mode of star formation. A direct method to quantify the physical and evolutionary states of these cores is through their density profiles. In hydrostatic equilibrium, the Bonnor-Ebert (B-E) sphere \citep{Bonnor1956} exhibits a Plummer-like density profile with a flat central region and an external power-law envelope. When the core collapses, the central flatten density radius shrinks, and the profile approaches a power-law. As the central density increases, the optical depth of the core grows. Whether the collapse is adiabatic or not, and whether the central H$_2$ dissociates, leads to the brief yet crucial First and Second Larson Cores \citep{Larson1972} between the prestellar and protostellar phases of low-mass star formation. During these phases, the collapse halts temporarily, and the density profile returns to a hydrostatic equilibrium resembling a B-E-like sphere. In the case of high-mass star formation, nuclear reactions are triggered at earlier stages, the presence of additional radiative pressure and the resulting Rayleigh-Taylor instability complicates the scale and existence of the First Larson Core \citep{Krumholz2005,Bhandare2018}, ultimately affecting the upper mass limit of high-mass stars. 

There have been observational studies obtaining the density profile of cores. The earliest studies in the 1990s employed millimeter and submillimeter dust emission data and found Plummer-like density profiles \citep{Andre1996,Ward-Thompson1994}. \citet{Alves2001,Hung2010} used dust extinction data to identify B68 as a standard isothermal B-E sphere. Currently, researchers have been using dust emission \citep{Beuther2002,Kirk2005,Palau2021}, dust extinction \citep{Alves2001}, and molecular tracer data \citep{Caselli2002,Tafalla2004} to extract density profiles of cores. They have unveiled stable and unstable B-E spheres, power-laws with various indices, as well as other forms such as polytropes and logotropes \citep{Pirogov2009}.

One of the effective methods to obtain the density profile of cores is the spectral energy distribution (SED) fitting of dust continuum data. This method provides 2D information on the core structure by solving modified blackbody functions that depend on the dust temperature, spectral index, and optical depth. It is widely used and has revealed various shapes of density profiles, such as power-laws and Plummers \citep{Launhardt2013}. The commonly used SED method usually assumes a constant dust temperature and spectral index along the line of sight, which can introduce significant biases in the derived density profile, particularly in cores with large temperature gradients \citep{Pagani2015}. These biases can lead to inaccurate representations of core structures, limiting our understanding of their formation and evolution.

To address these limitations, researchers have made efforts to improve the accuracy of dust emission modeling. Multi-component SED fitting allows dust components with different temperatures to coexist at the same position, providing a more detailed view of core structures (e.g., \citealt{Pagani2015,Miettinen2020}). More complex continuous multi-temperature mapping techniques have also been developed. The PPMAP method \citep{Marsh2015} decomposes multi-band dust continuum images into a finely binned grid of temperature and column density components, which helps resolve temperature variations along the line of sight and improves column density estimates. The inverse Abel transform \citep{Roy2014}, under the optically thin assumption, inverts dust emission intensities into 3D space for spherically symmetric sources, enabling the derivation of radial temperature distributions. Furthermore, for sources with complex 3D geometry, full radiative transfer modeling has been used to reconstruct dust temperature and density profiles (e.g., \citealt{Steinacker2005,Stamatellos2010}). Recent tools like the \textit{pandora} framework \citep{Scibelli2023} have wrapped the RADMC-3D code \citep{Dullemond2012} to achieve better grid-fit results. Despite these advancements, there remains a need for a detailed, layered core parameter fitting tool that accounts for radiative transfer while providing both accurate and efficient analysis of core properties.

A notable step in this direction is the COREFIT algorithm developed by \citet{Marsh2014}, which models Herschel observations of Taurus cores using spherically symmetric density and temperature profiles under the optically thin assumption. COREFIT combines parametric modeling with multi‑wavelength image fitting and demonstrated that incorporating spatial information significantly improves constraints on central temperatures, density indices, and core masses. Building on the same foundation, we introduce CARPP (Core Analysis via Radiative Transfer and Profile Parameters), which retains the strengths of parametric modeling while addressing several limitations of COREFIT. CARPP introduces four key enhancements: (i) the full radiative‑transfer equation is solved, avoiding the optically thin assumption; this is essential for cold cores and high-resolution observations, where the optically thin assumption can cause significant biases; (ii) the dust emissivity index $\beta$ is treated as a free parameter, enabling exploration of dust‑property variations; (iii) a more flexible temperature law is adopted, capturing the asymptotic behavior expected under external heating; and (iv) the code is publicly released to support systematic, reproducible studies.

CARPP is designed to balance physical realism with computational efficiency. Assuming spherical symmetry, it adopts parametric density and temperature profiles and integrates the radiative transfer equation along each line of sight through the spherical structure, enabling rapid generation of synthetic images across multiple wavelengths. This strategy preserves physical interpretability while remaining computationally efficient, offering a practical alternative to fully self‑consistent radiative‑transfer models. By fitting multi‑wavelength dust‑continuum data, CARPP recovers core properties with an accuracy that bridges the gap between simplified SED fitting and full radiative‑transfer simulations. CARPP is particularly advantageous for cores on the verge of protostar formation, where it provides detailed insights into their central temperatures and flat central density profiles, enabling direct comparisons with theoretical models. In Section~\ref{sec:CARPP}, we present the workflow of CARPP. Its robustness is tested in Section~\ref{sec:limit}. We apply CARPP to derive the properties of two real cores in Section~\ref{sec:real_core}. Section~\ref{sec:discuss} covers the discussion on CARPP's performance. The results are summarized in Section~\ref{sec:summary}.

\section{Algorithm Framework of CARPP}\label{sec:CARPP}

\subsection{Radiative Transfer Calculations}\label{sec:21}

CARPP treats cores as spherically symmetric, onion‑like structures and performs radiative transfer calculations along each line of sight through the concentric layers. For layer $i$ from a core's center, the radiation intensity is
\begin{equation}
I_{\nu}^i=
\begin{cases}
B_{\nu}^i(T^i)(1-{\rm e}^{-\tau_{\nu}^i}),& i=0 \\
I_{\nu}^{i-1}{\rm e}^{-\tau_{\nu}^i}+B_{\nu}^i(T^i)(1-{\rm e}^{-\tau_{\nu}^i}),& i>0
\end{cases}
\label{equ:1}
\end{equation}
where $B_\nu$ is calculated from Planck's radiation law and the optical depth is
\begin{equation}
\tau(\nu)=n_{\rm dust} \pi r_{\rm d}^2 Q(\nu)L,
\end{equation}
where the grain radius $r_{\rm d}$ is set to $0.1$ $\mu$m by default \citep{Mathis1977}. $L$ is the path length through this layer along the line of sight.
$Q(\nu)=Q_{350}\left(\frac{\lambda}{350 \space {\rm \mu m}}\right)^{-\beta}$ is the absorption efficiency. Note that $Q(\nu)$ is related to dust opacity $\kappa_\nu$ by 
\begin{equation}
Q(\nu)=\kappa_\nu\frac{4}{3}\rho_{\rm d}r_{\rm d}R_{\rm g-d},
\end{equation}
where we use $\rho_{\rm d}=3$ $\rm g/cm^3$ as the dust density \citep{Hildebrand1983} and $R_{\rm g-d}=100$ as the gas-to-dust mass ratio. 
$n_{\rm dust}$, the number density of dust grains, is given by
\begin{equation}
n_{\rm dust}=\frac{n_{\rm H_2}m_{\rm H_2}}{R_{\rm g-d}V_{\rm d}\rho_{\rm d}}.
\label{equ:4}
\end{equation}
$n_{\rm H_2}$ and $m_{\rm H_2}$ are the number density and mass of $H_2$. $V_{\rm d}=\frac{4}{3}\pi r_{\rm d}^3$ is the dust volume \citep{Hildebrand1983}.

The value of $Q_\nu$ is influenced by a variety of factors, including grain size distribution, shape, composition, and dust temperature \citep{Boudet2005,Meny2007,Ysard2019}. Current estimates of $Q_\nu$ remain subject to significant uncertainties, with reported $Q(350)$ values spanning a wide range from $0.5 \times 10^{-4}$ to $4 \times 10^{-4}$. For example, \citet{Pollack1994} obtained a value of $4.61 \times 10^{-5}$, \citet{Draine1984} reported $7.74 \times 10^{-5}$, 
and \citet{Goldsmith1997} measured $4.0 \times 10^{-4}$ (for a more comprehensive overview, see Table 5 in \citet{Goldsmith1997}). The observed flux density is directly proportional to the value of $Q(350)$. In CARPP, this parameter can be adjusted according to user requirements. By default, we use $Q_{350} = 1.36 \times 10^{-4}$ as suggested by \citet{Preibisch1993}, which is suitable for silicate grains covered by multiple types of ice.


\subsection{Density and Temperature Profile Assumptions}

CARPP assumes a generalized Plummer function for the core density profile, which we express throughout this work using the characteristic radius $r_0$,
\begin{equation}
  \rho=\frac{\rho_0 }{1+(r/r_0)^\alpha}
  \label{equ:rho_pf}
\end{equation}
where $\rho_0$ is central density and $\alpha$ is the density index. This functional form has been widely used in core modeling, including in the COREFIT algorithm \citep{Marsh2014}. The profile closely resembles core density profiles with various shapes, such as the B-E sphere (which approximates Plummer profile with $\alpha=2.4$) \citep{Tafalla2004}, and power-law density profiles (approximates Plummer profiles with smaller characteristic radii $r_0$). This approach ensures physically meaningful density distributions and simplifies computational complexity while offering flexibility to describe different core states. All the three parameters, $\rho_0$, $\alpha$, and $r_0$ are used as free parameters in CARPP.

For the temperature profiles of the core, CARPP assumes a continuous function given by
\begin{equation}
  T=T_1+\frac{T_0-T_1}{1+(r/r_{\rm t})^{\alpha _{\rm t}}}.
  \label{equ:T_pf}
\end{equation}
This temperature profile involves the characteristic radius $r_{\rm t}$, the index $\alpha _{\rm t}$, and two temperature parameters: $T_0$ at the central region and $T_1$ at infinity. It offers the flexibility to describe a wide range of temperature profiles, regardless of whether the core is cooler or warmer than its surrounding environment. In the default CARPP fit, we fit $T_0$, $T_1$, and $r_{\rm t}$. We fix $\alpha_t=2$ to reduce the number of free parameters. This value provides a smooth temperature transition suitable for most prestellar cores. Users can adjust $\alpha_t$ manually or include it in a grid search if steeper gradients are required.

\subsection{Workflow of the CARPP Algorithm}\label{sec:CARPP_workflow}

The main objective of CARPP is to obtain the density profiles, temperature profiles, and spectral indices of cores based on observed dust continuum images and spectral energy distribution data.

This tool comprises three key components:
\begin{enumerate}
\item Image preprocessing tool: This component extracts a square area centered on the core from the images, it uses input parameters and flux peaks as criteria for selection.
\item Forward generator: The forward generator generates simulated 2D images by employing a spherically symmetric radiative transfer model with predefined density and temperature profiles. 
\item Fitting tool: The fitting tool use the forward generator to fit the observed images. It extracts properties, including the density profile, temperature profile, and spectral index of the core, through a comparison between the simulated and observed images.
\end{enumerate}

\begin{figure}
  \centering
  \includegraphics[width=0.8\linewidth]{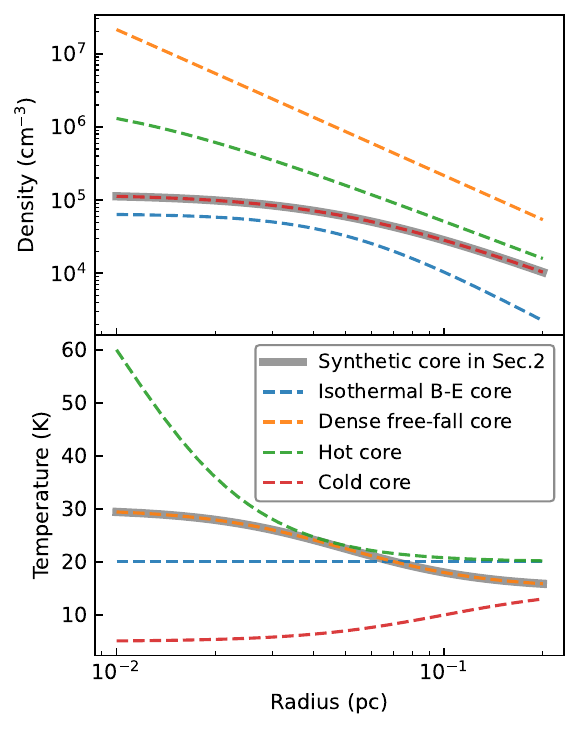}
  \caption{The density and temperature profiles of the synthetic core in Section~\ref{sec:CARPP} and other synthetic cores in Section~\ref{sec:4syn}. }
  \label{fig:calamid_profile}
\end{figure}

\begin{figure*}
  \centering
  \includegraphics[width=0.8\linewidth]{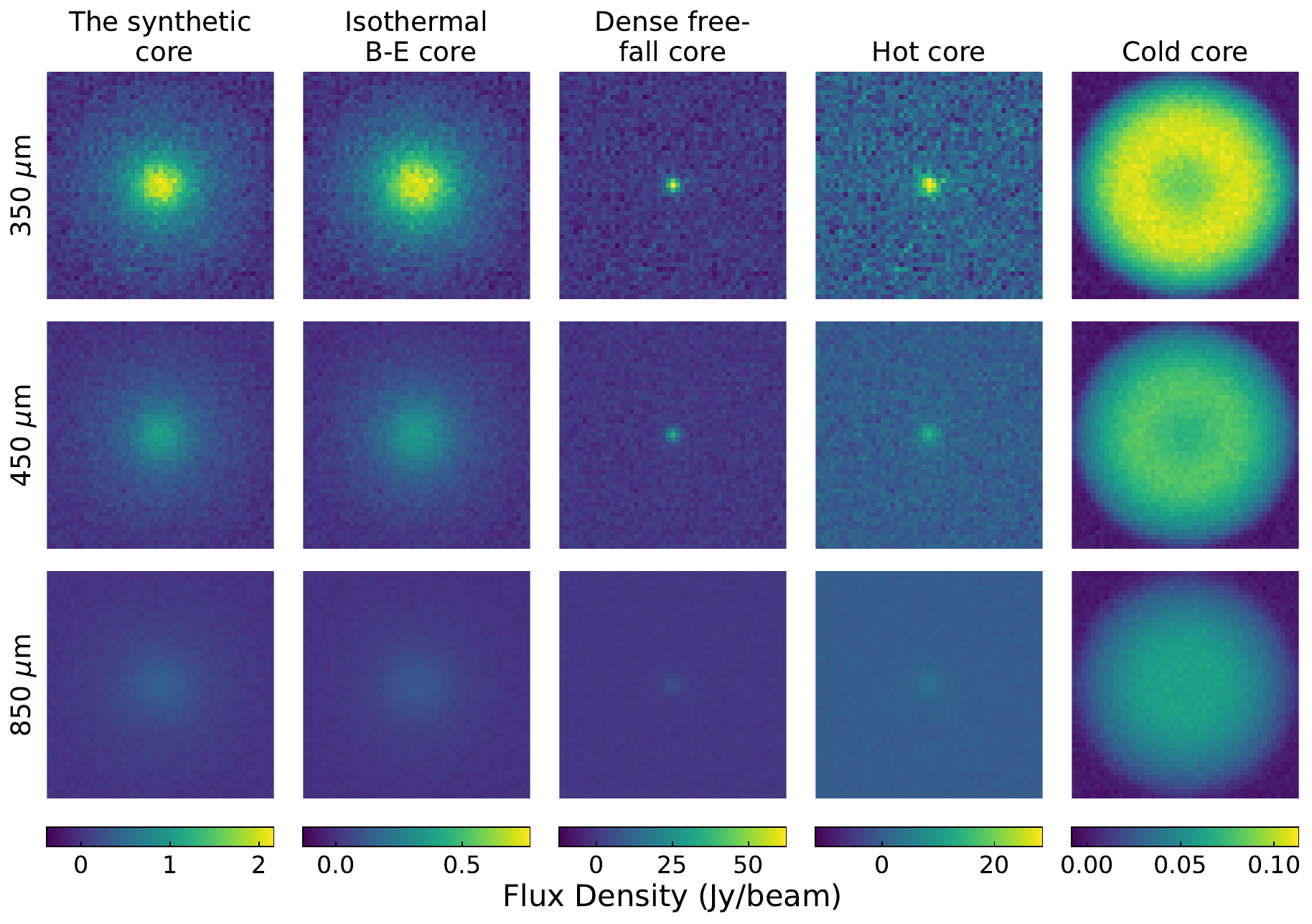}
  \caption{The dust continuum maps of the synthetic core in Section~\ref{sec:CARPP_workflow} and the four other cores in Section~\ref{sec:4syn}. The cores all have a diameter of 49 pixels, corresponding to 0.4 pc at $d=414$ pc. }
  \label{fig:calamid_fig}
\end{figure*}

We demonstrate the performance of CARPP using a synthetic core with $M=$ 30 $M_\odot$ and $R=$ 0.2 pc. The synthetic core has density and temperature profiles as in Equation~\ref{equ:rho_pf} and \ref{equ:T_pf}, and is shown in Figure~\ref{fig:calamid_profile}.The temperature index of the core is set to $\alpha _{\rm t} = 2$, and the other parameters can be found in Table~\ref{tab:result}. We put the core at the distance of Orion ($d=414$ pc) and calculate its dust emission maps at 350, 450, and 850 $\mu$m. To ensure the maps resemble authentic observational data, we employ a pixel scale of 4'' and convolve the maps to achieve resolutions of 8.5'', 8.5'', and 14'', respectively. We then add Gaussian noises of 0.05 times the peak fluxes to the emission maps. This noise level is chosen to reflect typical submillimeter observations of dense cores. For instance, in surveys using Herschel (e.g., the Herschel Gould Belt Survey; \citealt{Andre2010}) and SCUBA-2 data (such as the JCMT SCUBA-2 Gould Belt Survey; \citealt{Ward-Thompson2007, Kirk2016, Pattle2015}), typical prestellar and protostellar cores exhibit peak fluxes in the range of $\sim0.03$--$10$~Jy/beam with RMS noise levels ranging from a few mJy/beam up to $\sim50$~mJy/beam. This yields peak signal‑to‑noise ratios generally $\gtrsim 20$ for bright sources. The maps are shown in Figure~\ref{fig:calamid_fig}. 

\begin{table*}
\caption{The fitting results of CARPP and three alternative SED methods on synthetic cores.}
\label{tab:result}
\centering
\begin{tabular}{*{10}{c}}
\toprule
Fitting method & $\beta$ & $\rho_0$ & $\alpha$ & $r_0$ &$T_0$ & $T_1$ & $r_{\rm t}$ & reduced $\chi^2$ & Relative error\\
 & & (cm$^{-3}$) & & (0.01 pc) & (K) & (K) & (0.01 pc) & & (\%)\\
\hline
\multicolumn{10}{l}{\textbf{The synthetic core in Section~\ref{sec:CARPP}}} \\
Input properties & 1.60 & 1.20$\times 10^{5}$ & 1.70 & 5.00 & 30.00 & 15.00 & 5.00 & - & - \\
CARPP & 1.63 & 1.33$\times 10^{5}$ & 1.69 & 4.66 & 29.35 & 15.52 & 4.25 & 1.45 & 5.82 \\

350-850 $\mu$m fit & 1.50 & 1.16 $\times 10^{5}$ & 2.47 & 10.20 & 19.55 & 12.48 & 5.08 & - & 30.30 \\
850 $\mu$m estimate & - & 5.60 $\times 10^{5}$ & 1.83 & 3.02 & - & - & - &- & 1.38$\times 10^{2}$ \\
Abel inversion & 1.56 & 6.66 $\times 10^{4}$ & 3.01 & 8.96 & 26.76 & 10.00 & 15.25 & - & 64.63 \\

\hline
\multicolumn{10}{l}{\textbf{Other synthetic cores}} \\
\multicolumn{10}{l}{\textbf{Isothermal B-E core}} \\
Input properties & 2.00 & 6.50$\times 10^{4}$ & 2.40 & 5.00 & 20.00 & 20.00 & - & - & - \\
CARPP & 2.16 & 1.02 $\times 10^{5}$ & 2.28 & 4.49 & 16.55 & 17.30 & -1.18 & 1.36 & 18.48 \\

350-850 $\mu$m fit & 1.50 & 3.97 $\times 10^{4}$ & 2.57 & 5.79 & 24.13 & 24.63 & 11.67 & - & 21.77 \\
850 $\mu$m estimate & - & 1.03 $\times 10^{5}$ & 2.28 & 4.94 & - & - & - &- & 21.55 \\
Abel inversion & 1.90 & 3.50 $\times 10^{4}$ & 2.94 & 8.16 & 21.99 & 16.91 & 20.00 & - & 27.04 \\

\multicolumn{10}{l}{\textbf{Dense free-fall core}} \\
Input properties & 1.60 & 2.16$\times 10^{9}$ & 2.00 & 0.10 & 30.00 & 10.00 & 5.00 & - & - \\
CARPP & 1.54 & 2.50 $\times 10^{9}$ & 2.13 & 0.10 & 29.92 & 5.46 & 9.53 & 1.22 & 23.18 \\

350-850 $\mu$m fit & 1.50 & 2.77 $\times 10^{11}$ & 2.72 & 2.42 $\times 10^{-2}$ & 23.11 & 19.14 & 3.98 & - & 1.87$\times 10^{3}$ \\
850 $\mu$m estimate & - & 2.33 $\times 10^{12}$ & 6.34 & 0.36 & - & - & - &- & 3.61$\times 10^{4}$ \\
Abel inversion & 1.74 & 9.10 $\times 10^{6}$ & 3.71 & 6.48 & 49.71 & 16.98 & 0.10 & - & 9.73$\times 10^{2}$ \\

\multicolumn{10}{l}{\textbf{Hot core}} \\
Input properties & 1.60 & 2.61 $\times 10^{6}$& 2.00 & 1.00 & 100.00 & 20.00 & 1.00 & - & - \\
CARPP & 1.60 & 2.00 $\times 10^{6}$ & 1.81 & 1.24 & 96.82 & 13.65 & 1.14 & 1.15 & 17.55 \\

350-850 $\mu$m fit & 1.50 & 8.20 $\times 10^{4}$ & 1.45 & 5.80 & 175.37 & 26.01 & 1.14 & - & 1.04 $\times 10^{2}$ \\
850 $\mu$m estimate & - & 9.69 $\times 10^{11}$ & 3.31 & 5.91 $\times 10^{-2}$ & - & - & - &- & 1.24 $\times 10^{7}$ \\
Abel inversion & 1.70 & 3.49 $\times 10^{5}$ & 3.32 & 6.99 & 31.97 & 20.00 & 11.70 & - & 2.71 $\times 10^{2}$ \\

\multicolumn{10}{l}{\textbf{Cold core}} \\
Input properties & 1.60 & 1.20$\times 10^{5}$ & 1.70 & 5.00& 5.00 & 15.00 & 10.00 & - & - \\
CARPP & 1.62 & 1.82$\times 10^{5}$ & 1.58 & 3.40 & 4.51 & 14.62 & 9.01 & 1.09 & 16.33 \\

350-850 $\mu$m fit & 1.50 & 3.29 $\times 10^{4}$ & 2.84 & 11.95 & 9.84 & 15.55 & 13.62 & - & 60.12 \\
850 $\mu$m estimate & - & 1.71 $\times 10^{4}$ & 4.17 & 27.29 & - & - & - &- & 2.26$\times 10^{2}$ \\
Abel inversion & 1.65 & 1.39 $\times 10^{5}$ & 1.71 & 4.50 & 5.21 & 16.04 & 12.32 & - & 9.13 \\

\hline
\hline
\end{tabular}
\begin{flushleft}
\footnotesize
\textit{Notes.} 
`350--850 $\mu$m fit' is a multi‑wavelength modified blackbody fit assuming optically thin emission and a uniform temperature along each line of sight. `850 $\mu$m estimate' is a single-wavelength column density estimation that assumes a fixed spectral index $\beta = 2$ and a uniform dust temperature $T = 15$~K. `Abel inversion' is an Abel‑based profile reconstruction method that faithfully reproduces the approach of \citet{Roy2014}. 
The relative error is defined as the mean fractional deviation of the fitted parameters from their input values. For the 850\,$\mu$m estimates, only the three density parameters ($\rho_0$, $\alpha$, and $r_0$) are included. 
For the isothermal B-E core, the average excludes $r_{\rm t}$, resulting in six parameters being used.
For other cases, the average is taken over seven parameters ($\beta$, $\rho_0$, $\alpha$, $r_0$, $T_0$, $T_1$, and $r_{\rm t}$). 
\end{flushleft}
\end{table*}

The input data of CARPP consists of a set of observation images at different wavelengths, which can differ in resolution and pixel scale. As for the synthetic core, we use its dust emission maps at 350, 450, and 850 $\mu$m. If the location of the core's center is not specified, CARPP automatically identifies the pixel of peak intensity as the structural center. CARPP then aligns the maps to a common coordinate grid via subpixel registration and subtracts any user-specified background intensity (zero for the synthetic core).

CARPP uses a forward generator to convert physical profile parameters into convolved 2D model images for direct point-by-point comparison on the aligned coordinate grid. The forward generator treats the core as a multi-layer, onion-like spherically symmetric structure with density and temperature profiles given by Equations~\ref{equ:rho_pf} and \ref{equ:T_pf}, and solves the radiative transfer equation. The resulting synthetic maps are convolved with a 2D Gaussian beam to match the respective observational FWHM of each band. 

The forward generator is first used to execute a series of model grids, aims at identifying a set of favorable initial parameters for the fitting process. To reduce the number of grids and speed up the calculation, the central density $\rho_0$ is left unspecified. Instead, we temporarily treat the continuum image with the longest wavelength as optically transparent. Consequently, there are only six parameters in the model grids, and their values are presented in Table~\ref{tab:model_grid}. Regarding the synthetic core, the best parameter configuration is [$\beta$,$\alpha$,$r_0$,$T_0$,$T_1$,$r_{\rm t}$]=[1.5,2.0,12.5,30.0,20.0,12.5], yielding a reduced chi-squared value of 1.42. 

The best set of gridded parameters are processed to the fitting process. The fitting process employs the Powell’s fitting algorithm, an optimization algorithm used for multidimensional minimization. It is widely used for parameter estimation in nonlinear regression analysis. By default, seven parameters are involved in the fitting process: density parameters ($\rho _0$, $\alpha$, $r_0$), temperature parameters ($T_0$, $T_1$,$r_{\rm t}$), and the spectral index ($\beta$). Pre-setting certain parameter values allows for a reduction in the number of parameters considered. The fitting process works by minimizing the reduced chi-square between the regridded input images and the generated images. The chi-square is computed on a pixel-by-pixel basis and is weighted by the noise of each band $\sigma_n$. Defining $d_i$ as the distance of the pixel to the central center of the core, only pixels with $d_i$ less than the radius of the regridded input images are used. To ensure that all points on the logarithmic radial profile contribute equally, we weight each pixel by $a_i=1/d_i^2$, where for the central pixel ($d_i = 0$) we set $d_i$ to the pixel scale (the smallest non-zero distance among all pixels) to avoid division by zero. The weighting compensates for the geometric increase in pixel number with radius and suppresses the tendency of noisy outer annuli to dominate the chi-square. The weighting exponent can be changed by the user in practice.
\begin{equation}
  \chi^2=\sum_{n}^{N_{\rm band}}\sum_{i}^{N_{\rm pix}}\left[ \frac{a_i}{\overline{a_i}}\left( \frac{F_{{\rm obs},i,n}-F_{{\rm fit},i,n}}{\sigma_n} \right)^2\right]
  \label{equ:chi}
\end{equation}
The reduced chi-square is defined as $\chi^2_{\rm reduced}=\chi^2/K$, where the number of degrees of freedom $K=N_{\rm band}N_{\rm pix}-N_p$. Only pixels with a distance from the center pixel less than the total radius of the core are counted in $N_{\rm pix}$. $N_p$ denotes the number of fitting parameters, which is 7 by default. 

\begin{table*}
\caption{Parameters of the gridded models.}
\label{tab:model_grid}
\centering
\begin{tabular}{*{10}{c}}
\toprule
Parameter (unit) & gridded values\\
\hline
$\beta$ & 1.5, 1.6, 1.7, 1.8, 1.9, 2.0\\
$\alpha$ & 1.2, 1.4, 1.6, 1.8, 2.0, 2.2, 2.4\\
$r_0$ (layers) & 0.78, 1.56, 3.13, 6.25, 12.5, 25.0\\
$T_0$ (K) & 5.0, 10.0, 15.0, 20.0, 25.0, 30.0\\
$T_1$ (K) & 5.0, 10.0, 15.0, 20.0, 25.0, 30.0\\
$r_{\rm t}$ (layers) & 0.78, 1.56, 3.13, 6.25, 12.5, 25.0\\
\hline
\hline
\end{tabular}
\begin{flushleft}
\footnotesize
\textit{Notes.} 
The term `layer' refers to the number of shell in the onion‑like, spherically symmetric structure used in the forward generator. The core is discretized into 50 layers by default, counted outward from the center.
\end{flushleft}
\end{table*}

Subsequently, the set of parameters resulting in the minimum reduced chi-square is considered as the fitting outcome. In the case of the synthetic core, we present the fitting results in Table~\ref{tab:result}. CARPP exhibits a high level of accuracy in obtaining information about the synthetic core. The derived parameters align closely with the input values, with a reduced chi-square value of $1.45$ and an average relative error of 5.82\% for the seven parameters.

\section{Robustness of CARPP} \label{sec:limit}

\subsection{CARPP Performance on Synthetic Cores}\label{sec:4syn}

To assess the fitting performance of CARPP across different cases, we fitted four additional synthetic cores whose parameters were chosen to be as similar as possible to the synthetic core in Section~\ref{sec:CARPP_workflow} while preserving distinct structural and thermal properties. All cores were placed at the same distances and imaged at the same resolutions as the reference synthetic core, their synthetic observations are shown in Figure~\ref{fig:calamid_fig}. Gaussian noise equal to 0.05 times the peak flux was added to every core except the `Cold core', for which a lower noise level of 0.025 times the peak flux was used because of its faint central emission, corresponding to a core observed with deeper integration and thus a signal‑to‑noise ratio of 40. The input parameters and the fitted results are listed in Table~\ref{tab:result}.

We also compared CARPP with three alternative methods. The first method, `350--850 $\mu$m fit', is a multi-wavelength 2D SED fitting method that assumes optically thin dust emission and a uniform temperature along each line of sight. We first convolved the maps to the 850 $\mu$m resolution. Simple pixel‑by‑pixel 2D SED analyses like this are strongly affected by noise in low signal‑to‑noise regions, and many methods that employ regularization or convolution have been developed to suppress noise \citep{Kelly2012,Marsh2015}. For our spherically symmetric synthetic cores, to reduce the influence of noise we extracted the 1D radial average intensity profiles across the three bands at 350, 450, and 850$\mu$m and performed radial SED fitting. This yields the column density and temperature profiles, together with a single global value of $\beta$. We then inverted the column density profiles using Equations~\ref{equ:rho_pf} to obtain the volume density, and approximated the temperature profiles using Equations~\ref{equ:T_pf}. 

The second method, `850 $\mu$m estimate' is a single-wavelength column density estimation that assumes a fixed spectral index $\beta = 2$ \citep{Planck2011,Montillaud2015} and a uniform dust temperature $T = 15$~K. The column density is calculated directly from the mock 850 $\mu$m intensity map using Equation~\ref{equ:normal_sed}. We then extracted its radial column density profile and inverted it to obtain the parameters ($\rho_0$, $r_0$, and $\alpha$) of the volume density profile. In both the methods above, the column density $N$(H$_2$) is calculated via Equations \ref{equ:1}–\ref{equ:4}, specifically
\begin{equation}
N({\rm H_2})=\frac{I_{\nu} D^2}{\kappa_\nu B_\nu(T) m_{\rm H_2}},
\label{equ:normal_sed}
\end{equation}
where $D$ is the distance to the source. 

The third method, `Abel inversion', is an Abel‑based profile reconstruction method that faithfully reproduces the approach of \citet{Roy2014}. First, all the three continuum maps at 350, 450, and 850$\mu$m are convolved to the lowest resolution among the bands. Then the 1D radial average intensity distribution is obtained for each wavelength and smoothed with a sixth‑order even polynomial. Under the assumption of spherical symmetry, the Abel analytical transform is applied to directly invert the volume emission coefficient at each band as a function of physical radius $r$. Finally, a single‑temperature modified blackbody fit is performed at each radius to derive the volume density and temperature profiles. The resulting radial profiles are then fitted using the parametric forms in Equations~\ref{equ:rho_pf} and~\ref{equ:T_pf}. 

CARPP performs consistently well across all synthetic cores, yielding small relative errors and acceptable reduced $\chi^2$ values. All methods perform reasonably on the isothermal Bonnor--Ebert core. In addition, the 350--850\,$\mu$m fit and the Abel inversion method exhibit good performance on the synthetic core described in Section~\ref{sec:CARPP} and on the cold core. The primary physical difference between CARPP and the three alternative methods lies in the handling of dust optical depth. All three alternative methods fundamentally rely on the optically thin assumption. In relatively diffuse environments, this assumption is reasonably valid, allowing even the Abel inversion method to achieve a low relative error of 9.13\% for the cold core. However, when applied to dense, highly compressed environments such as the dense free-fall core ($\rho_0 = 2.16\times10^9\,\rm cm^{-3}$), where the central line-of-sight optical depths have $\tau >1$ at 350 and 450\,$\mu$m, these methods fail catastrophically. In such dense regions, the optically thin assumption leads to severe errors in the derived density parameters, particularly $\rho_0$ and $r_0$. In contrast, CARPP solves the full, non-linear radiative transfer along each line of sight without any optically thin simplification, thereby successfully reproducing the properties of even the highly optically thick dense free-fall core with a moderate relative error of 23.18\%.

A second distinction is the handling of multi-temperature components along the line of sight. The 350--850\,$\mu$m fit and the 850\,$\mu$m estimate assume a uniform temperature along each line of sight, which leads to an underestimation of the temperature gradient in several cores. For the 350--850\,$\mu$m fit, the bias in temperature also leads to a large error in $\beta$ due to the $\beta$--$T$ degeneracy,This degeneracy drives the fitted emissivity index toward the lower bound of the allowed range (in our fits, $\beta = 1.5$ for all cores, regardless of the true input value). For the 850\,$\mu$m estimate, the strong dependence on the assumed temperature and spectral index makes it reliable only under very restrictive conditions. By contrast, both CARPP and the Abel inversion method allow different radii within the spherically symmetric core to have different temperatures, thus providing a more accurate representation of the radial temperature profile.

Furthermore, the treatment of beam convolution further distinguishes CARPP from other methods. CARPP incorporates 2D beam convolution inside its forward generator, preserving the native resolution of each input image. It allows high-spatial-resolution short-wavelength observations to compensate for poorer resolution at longer wavelengths, helping the reconstruction of central core properties (such as central density $\rho_0$ and characteristic flat radius $r_{0}$). In contrast, both the Abel inversion and the 350--850\,$\mu$m fit require all maps to be convolved to the lowest common resolution (the 850\,$\mu$m map, which has a beam size of 14\arcsec) before analysis. This resolution degradation makes it more difficult for these methods to recover the fine structural details in the central regions of cores with concentrated intensity distributions, such as the dense free-fall core and the hot core.

\subsection{Degeneracy Analysis}
\begin{figure*}
  \centering
  \includegraphics[width=\linewidth]{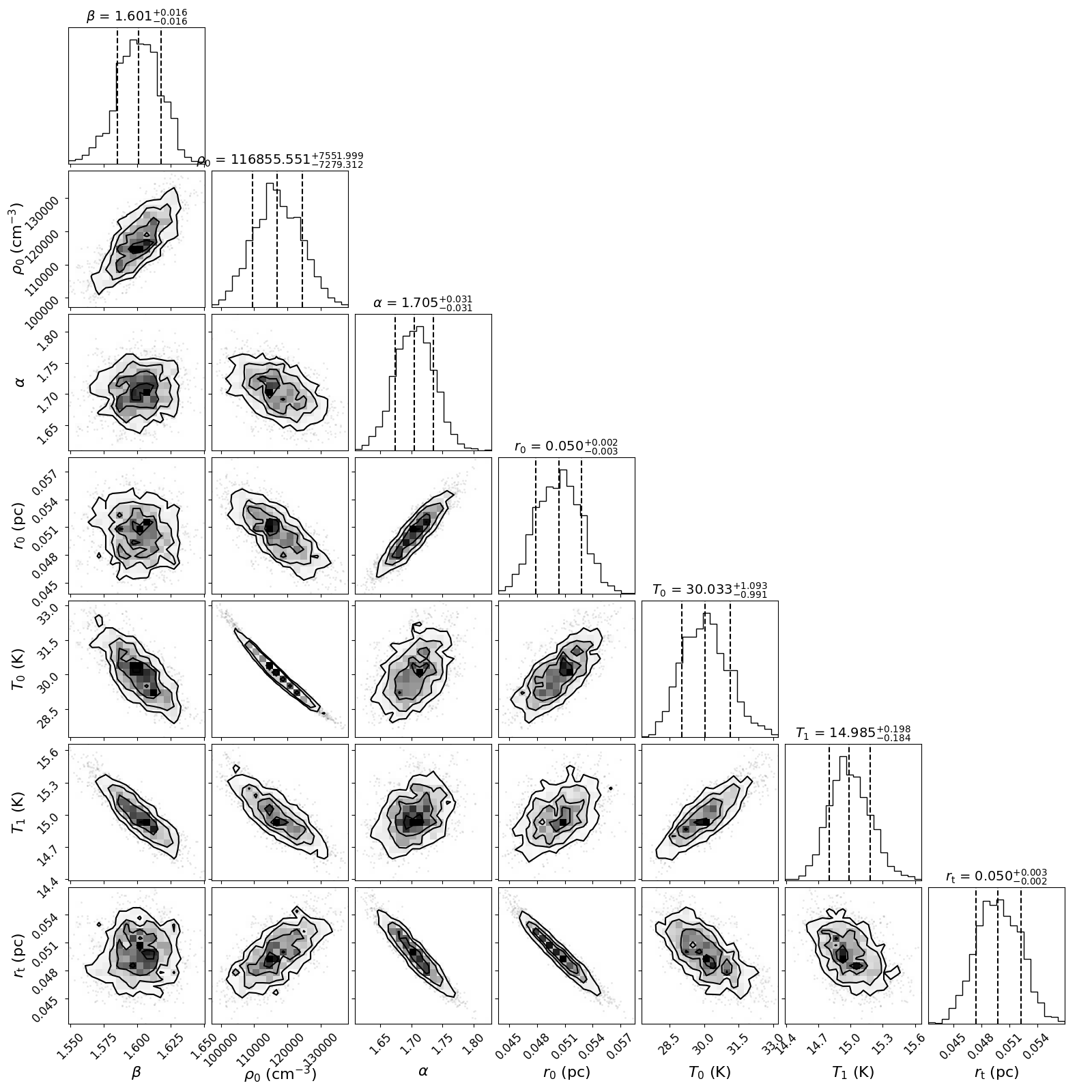}
  \caption{Corner plot showing the posterior probability distributions for the seven parameters, marked with $1\sigma$ standard deviations. The plot is derived from an MCMC run on noiseless data, illustrating the intrinsic degeneracies within the model.}
  \label{fig:corner}
\end{figure*}

In CARPP, the interplay of the physical parameterization, the convolution with different FWHMs, and the effects of observational noise can lead to significant degeneracies among the seven fitted parameters. To acknowledge these degeneracies, we performed a Markov Chain Monte Carlo (MCMC) analysis using the highest-resolution noiseless core used in Figure~\ref{fig:reso_sigma_err}. This core has the same physical parameters as the synthetic core described in Section~\ref{sec:CARPP_workflow}. For the MCMC analysis, we adopted uniform priors over physically motivated ranges (core mass $M$ in [1, 100] M$_\odot$, $\beta$ in [1.0,2.5], $\alpha$ in [1.0, 3.0], $r_0$ in [0.01, 0.2] pc, $T_0$ in [5, 60] K, $T_1$ in [5, 30] K, and $r_{\rm t}$ in [0.01, 0.2] pc). The range of central density $\rho_0$ is derived from $M$ and the density profile. We used 32 walkers, ran 2000 steps and discarded the first 1000 steps as burn-in. The resulting posterior probability distributions, shown as a corner plot in Figure~\ref{fig:corner}, reveal several well-defined degeneracy structures.

The most prominent degeneracy is the anti-correlation between the central density ($\rho_0$) and the central temperature ($T_0$). The observed dust emission flux density is proportional to the product of the column density and the Planck function integrated along the line of sight. To maintain a constant flux at a given wavelength, a slight decrease in $T_0$ must be compensated by a significant, non-linear increase in $\rho_0$ to offset the exponential drop in the Planck function. This trade-off results in a narrow, diagonal negative correlation in the posterior. The plot also demonstrates the well-known $\beta$--$T$ degeneracy. When the fitted dust emissivity index $\beta$ becomes shallower, the model predicts relatively less emission at shorter wavelengths. To recover the observed flux ratios between short (e.g., 350 $\mu$m) and long (e.g., 850 $\mu$m) wavelengths, the fitting algorithm is forced to raise $T_0$. This behavior manifests as a strong positive correlation in the $\beta$--$T_0$ posterior. A significant positive correlation is observed between the density power-law index ($\alpha$) and the characteristic flattening radius ($r_0$). For the Plummer-like density profile, an increase in $r_0$ is accompanied by a steeper $\alpha$ to preserve the total mass and the flux distribution, particularly in the outer envelope. This compensation creates the observed diagonal positive correlation. The temperature characteristic radius ($r_{\rm t}$) shows negative correlations with both $\alpha$ and $r_0$, suggesting that the scale of the temperature gradient is coupled to the density structure. For instance, a more extended temperature plateau (larger $r_{\rm t}$) can mimic the effect of a different density gradient, leading to these anti-correlations.

\subsection{Effects of Noise and Resolution}

\begin{figure}
  \centering
  \includegraphics[width=\linewidth]{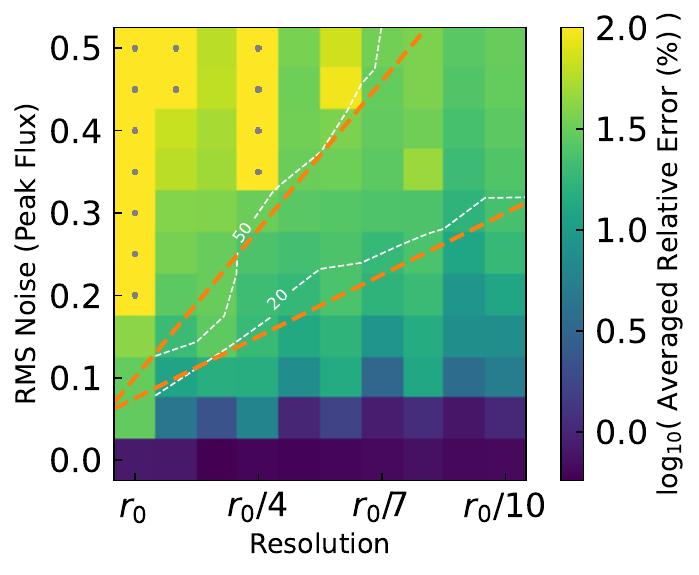}
  \caption{
  The averaged relative error of the seven parameters under different noise and resolution levels. The `+' (`-') marker indicates that the error is higher (lower) than the value indicated by the colorbar. The contours for 20\% and 50\% average relative error are shown in white, and the orange dashed lines represent the approximate visual estimates to these contours given by Equations~\ref{equ:noise_resolu_20} and \ref{equ:noise_resolu_50}, respectively. Below each orange line, the averaged relative error of the seven fitted parameters remains smaller than the corresponding threshold (20\% or 50\%).}
  \label{fig:reso_sigma_err}
\end{figure}

Excessive noise or insufficient resolution can result in missing information and increased uncertainty. To demonstrate how they affect the performance of CARPP, we add different levels of random noise to the synthetic core in Section~\ref{sec:CARPP} while maintaining its physical parameters, and sample it at different resolutions. During each trial, we obtained three images at 350, 450, and 850 $\mu$m, all with the same resolutions. For our synthetic core, the resolution of $r_0-r_0/10$ in Figure~\ref{fig:reso_sigma_err} corresponds to $0.8-0.08$ times the FWHM of the core's flux map, and $0.25-0.025$ times the full radius of the core. In all cases, the pixel scale is set to half the FWHM, ensuring Nyquist sampling and mimicking real observational conditions. To generate reliable results, we repeated this procedure 20 times for each noise and resolution level by regenerating random Gaussian noise. Subsequently, we applied the CARPP algorithm to fit the core. The fitting accuracy are presented with averaged relative error $\frac{1}{7} \sum_{p} |\frac{p_{\rm fit}-p_{\rm real}}{p_{\rm real}}|$, where $p$ represents the seven parameters in CARPP. The results in Figure~\ref{fig:reso_sigma_err} indicate that lower noise levels and higher resolutions are advantageous for achieving more precise fits with CARPP. Through visual assessment, we determine that when the relation between resolution and RMS noise satisfies 
\begin{equation}
   \frac{\rm RMS \,\, noise}{[\rm peak \,\, flux]} < 0.025\times \frac{[r_0]}{\rm resolution} +0.05,
  \label{equ:noise_resolu_20}
\end{equation}
the averaged relative error of the seven parameters falls below 20\%. And when
\begin{equation}
  \frac{\rm RMS \,\, noise}{[\rm peak \,\, flux]} < 0.06\times \frac{[r_0]}{\rm resolution} +0.04,
  \label{equ:noise_resolu_50}
\end{equation}
the averaged relative error of the seven parameters falls below 50\%. 
Furthermore, we extended our tests to include synthetic cores with the same density profiles but different temperature profiles. The above condition remains approximately valid: when the prescribed relations between resolution and RMS noise is satisfied, the averaged relative error of the fitted parameters generally stays below 20\% (or 50\%).

\begin{figure}
  \centering
  \includegraphics[width=\linewidth]{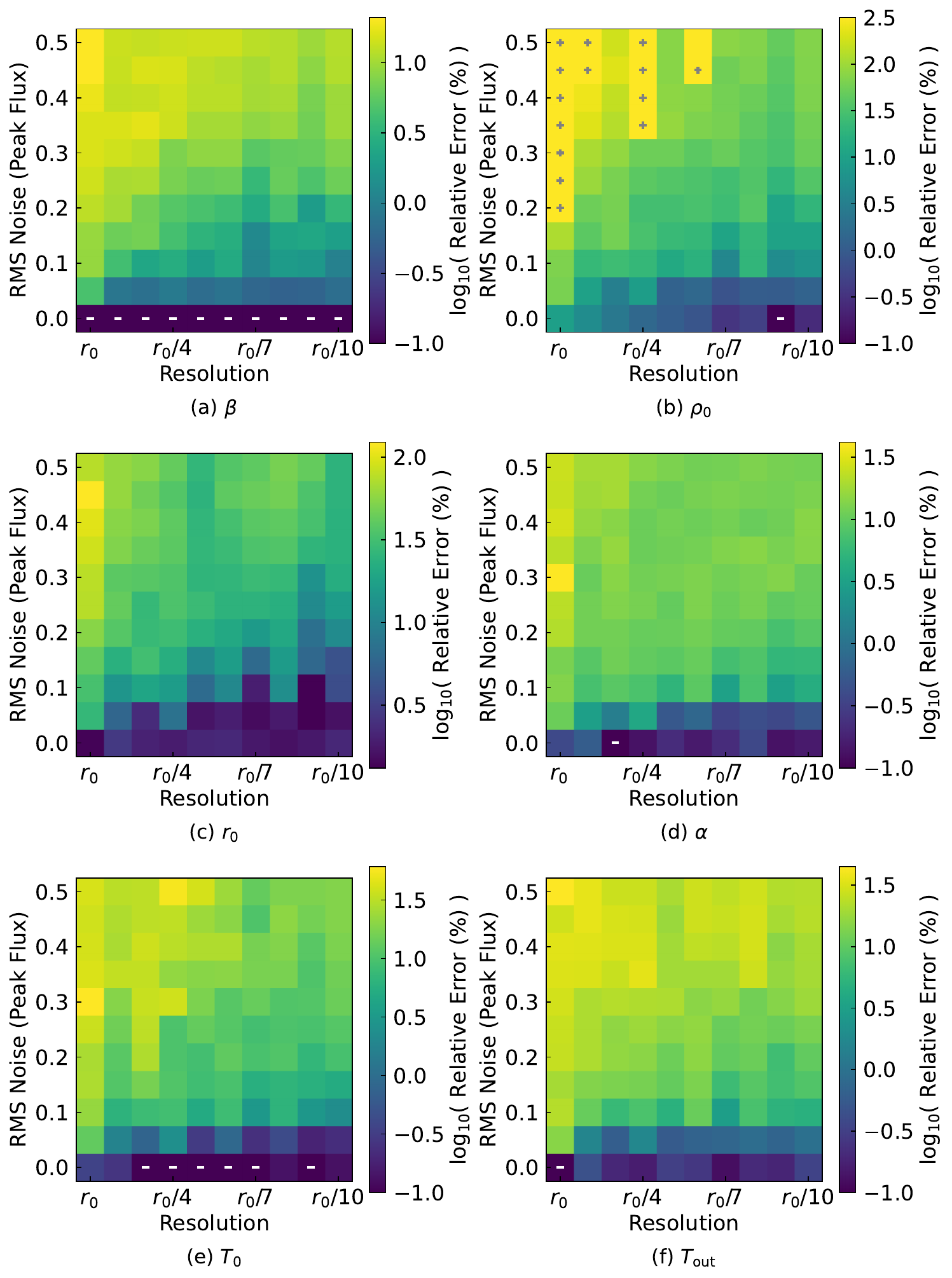}
  \caption{The relative error of the parameters at different noise and resolution levels. The makers are same as Figure~\ref{fig:reso_sigma_err}.}
  \label{fig:reso_sigma_4}
\end{figure}

In Figure~\ref{fig:reso_sigma_4}, we examine the errors of the fitted parameters at different noise and resolutions. To make the results easier to compare with the input values, instead of using that of $r_{\rm t}$ and $T_1$, we calculate the temperature at the full radius of the core $T_{\rm out}$ and obtain its relative error. All the parameters have larger errors when noise and resolution are poor. The estimation of $\rho_0$ proves to be most sensitive to both noise and resolution levels, detailed investigation reveals that under conditions of insufficient noise and resolution, CARPP tends to overestimate $\rho_0$, regardless of the core's temperature profile. In contrast, the fitting results for $\beta$ exhibit minimal susceptibility to variations in resolution and random noise, when considering the parameter settings represented in the Figure, the error remains within an acceptable margin of approximately 10\%. 

\subsection{Radius of the core} 
\label{sec:core_radius}

\begin{figure}
\centering
\includegraphics[width=\linewidth]{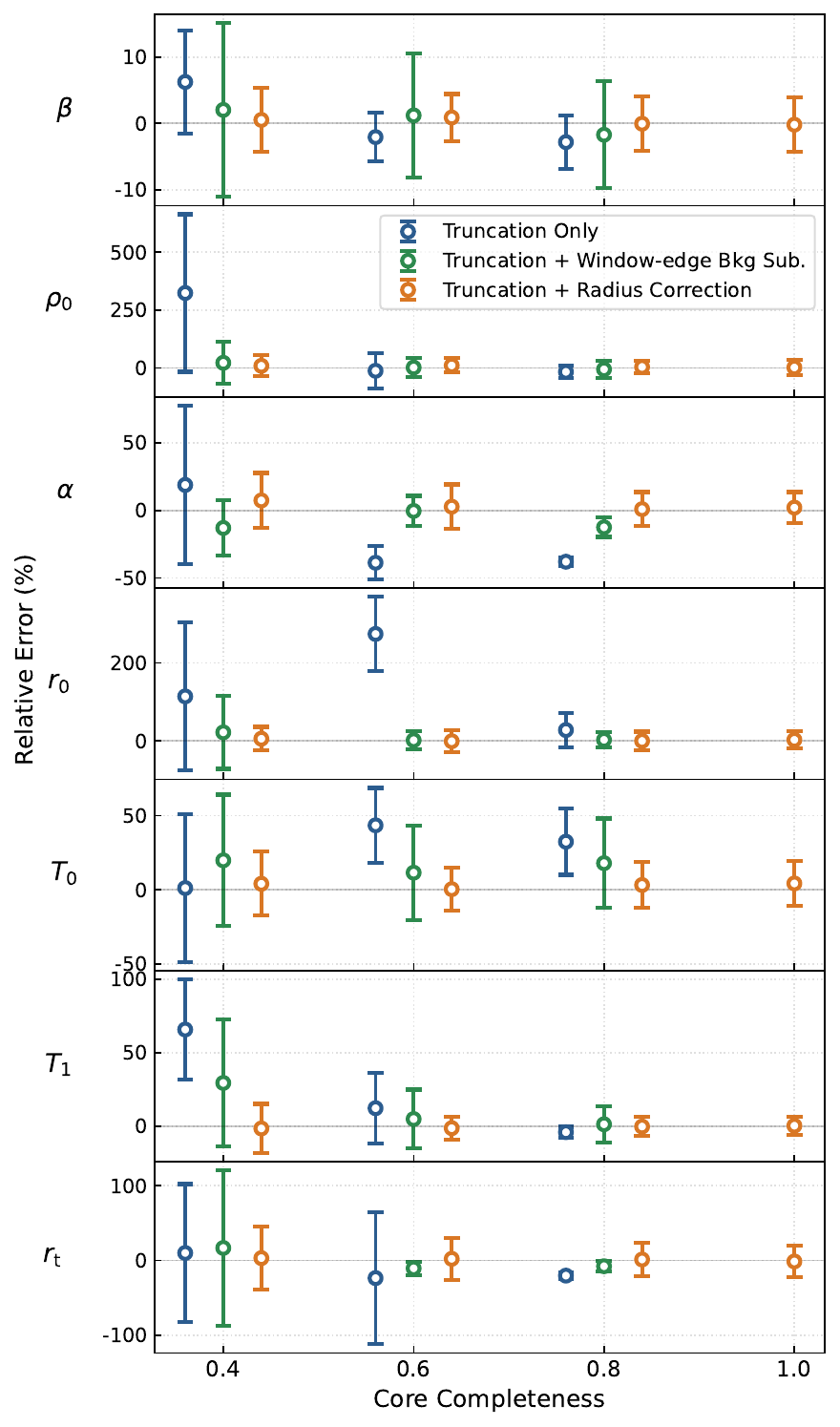}
\caption{Relative errors in the recovered parameters as a function of core completeness ($R_{\text{use}}/R$). Three scenarios are compared: spatial truncation only (blue), truncation accompanied by a window-edge background subtraction that modifies interior pixel values (green), and truncation with the true physical radius supplied as $R_{\text{est}}=R$ (orange). At a core completeness of 1.0, all three scenarios physically converge to the same un-truncated state. Points and errorbars denote the mean and $1\sigma$ standard deviation calculated from 100 realizations.}
\label{fig:bkg_Rest}
\end{figure}

CARPP requires the total radius $R$ of the core to set the outer boundary for the radiative-transfer calculation. In practice, estimating the full physical extent of a core from observations is nontrivial: a limited field of view, finite sensitivity, background gradients, and nearby sources can all truncate or confuse the apparent emission. To quantify how the radius of the used data ($R_{\text{use}}$) and its coupling with background subtraction affect the parameter recovery, we performed a controlled numerical experiment using the synthetic core from Section~\ref{sec:CARPP}. For each chosen core completeness ($R_{\text{use}}/R \in [0.4, 1.0]$), we considered three observational scenarios:
\begin{enumerate}
\item Truncation Only: The maps are simply truncated to $R_{\text{use}}$, and the pixel values inside remain unchanged. This serves as the baseline to isolate the effect of spatial truncation.
\item Truncation + Window-edge Bkg Sub.: The maps are truncated to $R_{\text{use}}$, and the emission at the new window edge is considered as the background level and is subtracted from the entire map. In this case, changing the assumed core size directly alters the interior pixel values, mimicking a common source of error in real observations where the background is misidentified due to a limited field of view.
\item Truncation + Radius Correction: The maps are truncated to $R_{\text{use}}$, but we explicitly inform CARPP of the true physical boundary by setting the estimated radius $R_{\text{est}}$ to the true core radius $R$.
\end{enumerate}
We repeated each experiment 100 times with independent realizations of Gaussian noise. The resulting mean relative errors and their $1\sigma$ standard deviations are presented in Figure~\ref{fig:bkg_Rest}.

When the core is truncated, the reduction in available spatial information leads to less accurate parameter estimates and larger uncertainties. This degradation is most pronounced in the ``Truncation only'' scenario, particularly under severe data incompleteness, where $\rho_0$ and $r_0$ can experience relative errors exploding by up to several hundred percent. This is because the radiative-transfer model attempts to compensate for the missing external emissions by severely distorting the global density distribution. Counterintuitively, introducing a window-edge background subtraction alongside truncation relieves these systematic biases. This improvement arises because subtracting the edge intensity removes some of the low-density envelope emission from the data, making the optimization algorithm focus mainly on restoring the characteristics of the center rather than compensating for missing external boundaries. Concurrently, supplying the true physical radius as $R_{\text{est}} = R$ yields the most robust parameter recovery when dealing with incomplete data. Therefore, to achieve better fitting results, it is highly recommended to acquire observational data that capture the full physical extent of the core to ensure a maximum $R_{\text{use}}$. When wide-field mapping data are unavailable and severe truncation is unavoidable, establishing an accurate constraint on the total core radius through ancillary data (e.g., low-resolution wide-field continuum maps or extinction maps) proves useful. Meticulous background evaluation and selection are also critical, as the coupling between spatial boundaries and local background subtraction has a profound influence on the recovered physical parameters of the core.

\section{Application of CARPP to Real Cores}\label{sec:real_core}

\subsection{TMC-1C}

We applied CARPP to observations of the prestellar cores TMC-1C and Ori2-2 to demonstrate its performance on real-world data. TMC-1C is a well-studied prestellar core in the Taurus molecular cloud ($d \approx 140$~pc). \citet{Crapsi2005} derived a central temperature of 4.4~K using N$_2$H$^+$ data, while multiple tracers (N$_2$H$^+$, CCS, and NH$_3$) indicate clear infall signatures \citep{Schnee2007,Koley2022}. Previous SED fitting by \citet{Schnee2005} suggested a central density of $10^{6-7}$~cm$^{-3}$ with a temperature drop towards the centre ($T_0 < 10$~K). Using broader wavelength coverage, \citet{Schnee2010} updated the spectral index to $\beta \approx 1.8$--$2.0$ and reported a line-of-sight average temperature towards the center of the core of 7~K.

\begin{figure}
  \centering
  \includegraphics[width=\linewidth]{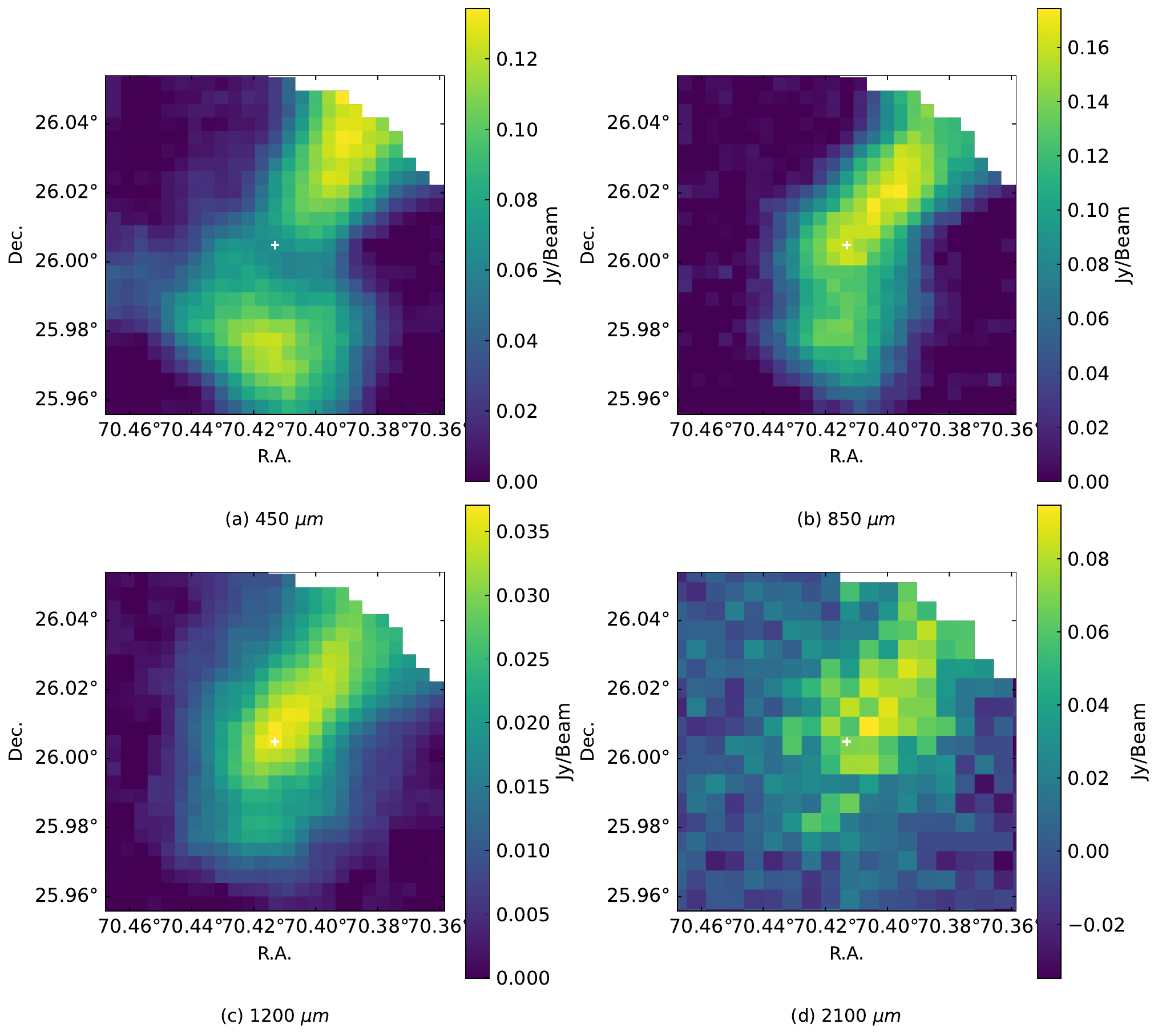}
  \caption{The dust continuum maps of TMC-1C. The maps show the central 0.12 pc region. }
  \label{fig:TMC-1Cfig}
\end{figure}

\begin{figure}
  \centering
  \includegraphics[width=0.8\linewidth]{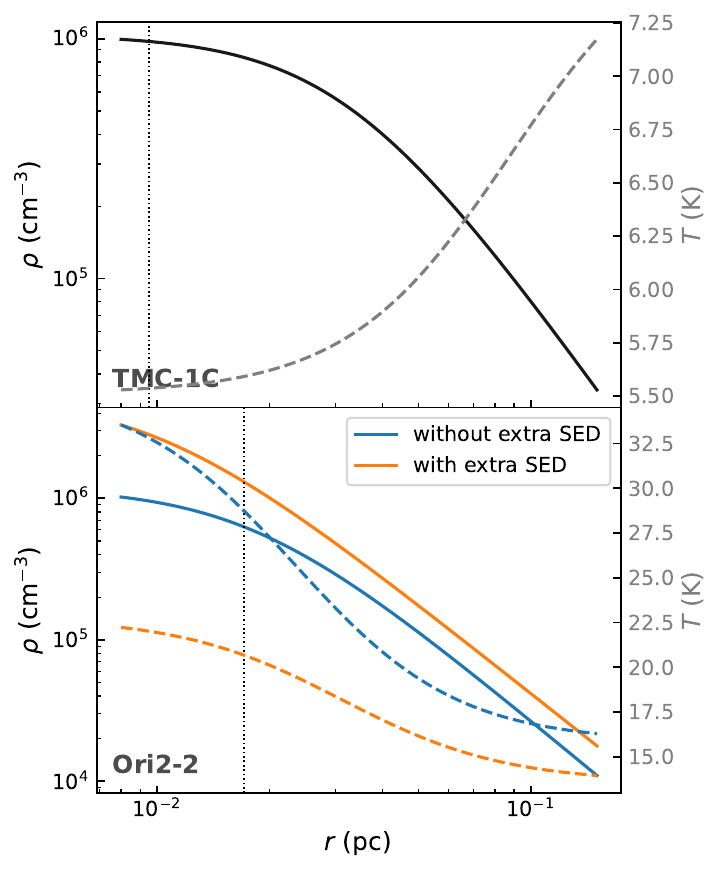}
  \caption{The fitted density and temperature profiles of TMC-1C and Ori2-2. The density profiles are in solid lines and the temperature profiles are in dashed lines. The vertical lines mark the best resolutions of the input data, which are 14$''$ and 8.5$''$, respectively.}
  \label{fig:nffig}
\end{figure}

We ran CARPP using dust continuum maps at 450, 850~$\mu$m (JCMT/SCUBA), 1200~$\mu$m (IRAM 30-m/MAMBO), and 2100~$\mu$m (CSO/Bolocam). The beam sizes of these instruments were originally 7.5$''$, 14.0$''$, 10.7$''$, and 60.0$''$. The 450, 850, and 1200~$\mu$m data are taken from \citet{Schnee2007_0}. In that work, the original observations were smoothed and regridded to a common resolution of 14$''$. We directly used the final map products in that work, with noise levels measured at this resolution of approximately 3.13, 6.30, and 0.93~mJy/beam, respectively. The CSO 2100~$\mu$m data retain a beam size of 60.0$''$ and has an RMS noise level of 14.90~mJy/beam. These values correspond to peak signal-to-noise ratios of approximately 42, 28, 40, and 6. To isolate the main core from two adjacent components to the northwest, we applied a mask as shown in Fig.~\ref{fig:TMC-1Cfig}. The 1200~$\mu$m map, being relatively optically thin and having high image quality, was used to define the core extent ($R_{\rm out} = 0.12$~pc). We measured the background and RMS noise levels from regions outside 0.12 pc and used data within 0.12 pc of the center for fitting. We then run the model grids and find the best parameter configuration is [$\beta$, $\alpha$, $r_0$, $T_0$, $T_1$, $r_{\rm t}$]=[1.9, 2.2, 3.125, 5.0, 10.0, 12.5]. The configuration indicates that TMC-1C has a high spectral index, consistent with the result of \citet{Schnee2007} and \citet{Schnee2010}. The final fitting results are listed in Table~\ref{tab:real}. The relatively large chi-squared value of the fit is attributed to the core's complex morphology and its elongation in the northwest-southeast direction. The density and temperature profiles of TMC-1C are shown in Figure~\ref{fig:nffig}. The density profile exhibits a distinct central flatten density region. When regarding TMC-1C as a B-E sphere, the ratio of the central to surface density is close to the critical value for hydrostatic equilibrium, which has density contrast $\rho_0/\rho_R \approx 14.1$ \citep{Bonnor1956,McCrea1957}. The central density at $10^6$ cm$^{-3}$ is in agreement with the detailed 2D SED fitting reported by \citet{Schnee2010}. The temperature profile is characterized by a cold central region surrounded by warmer outskirts, with a fitted core temperature $T_0=$5.51 K, similar to that derived from N$_2$H$^+$ observations \citep{Crapsi2005}. These results collectively support the interpretation that TMC-1C is an prestellar optically thick cold core. 

\begin{table*}
\caption{The fitting results of CARPP on TMC-1C and Ori2-2.}
\label{tab:real}
\centering
\begin{tabular}{*{9}{c}}
\toprule
Name & $\beta$ & $\rho_0$ & $\alpha$ & $r_0$ & $T_0$ & $T_1$ & $r_{\rm t}$ & reduced $\chi^2$\\
 & & (cm$^{-3}$) & & (0.01 pc) & (K) & (K) & (0.01 pc) & \\
\hline
\multicolumn{9}{l}{\textbf{TMC-1C}} \\
TMC-1C & 2.00 & 1.04$\times 10^{6}$ & 2.20 & 3.23 & 5.51 & 7.74 & 8.76 & 26.23 \\
\hline
\multicolumn{9}{l}{\textbf{Ori2-2}} \\
without extra SED & 1.50 & 1.20$\times 10^{6}$ & 2.20 & 1.79 & 35.60 & 15.82 & 2.34 & 1.23 \\
with extra SED & 1.98 & 5.41$\times 10^{6}$ & 2.10 & 1.00 & 22.79 & 13.57 & 3.12 & 1.89 \\
\hline
\hline
\end{tabular}
\end{table*}

\subsection{Ori2-2}

Ori2-2 is a dense core in Orion-A at a distance of 414 pc. The name Ori2-2 is quoted from \citet{Li2007}, who labeled the core as a quiescent core due to the absence of IRAS point sources and no association with molecular outflows. Using dust continuum data, they found the core has a mass of 33 $M_\odot$ with a radius of 0.029 pc. However, \citet{Tatematsu2010} classified it as a star-forming core because it has Spitzer 24 $\mu$m emissions. \citet{Stanke2002} suggested it could be a driver source for a candidate H$_2$ flow. \citet{Tatematsu2008} obtained a temperature of 9.2 K at the center of this core using N$_2$H$^+$ and HC$_3$N observations.

\begin{figure}
  \centering
  \includegraphics[width=\linewidth]{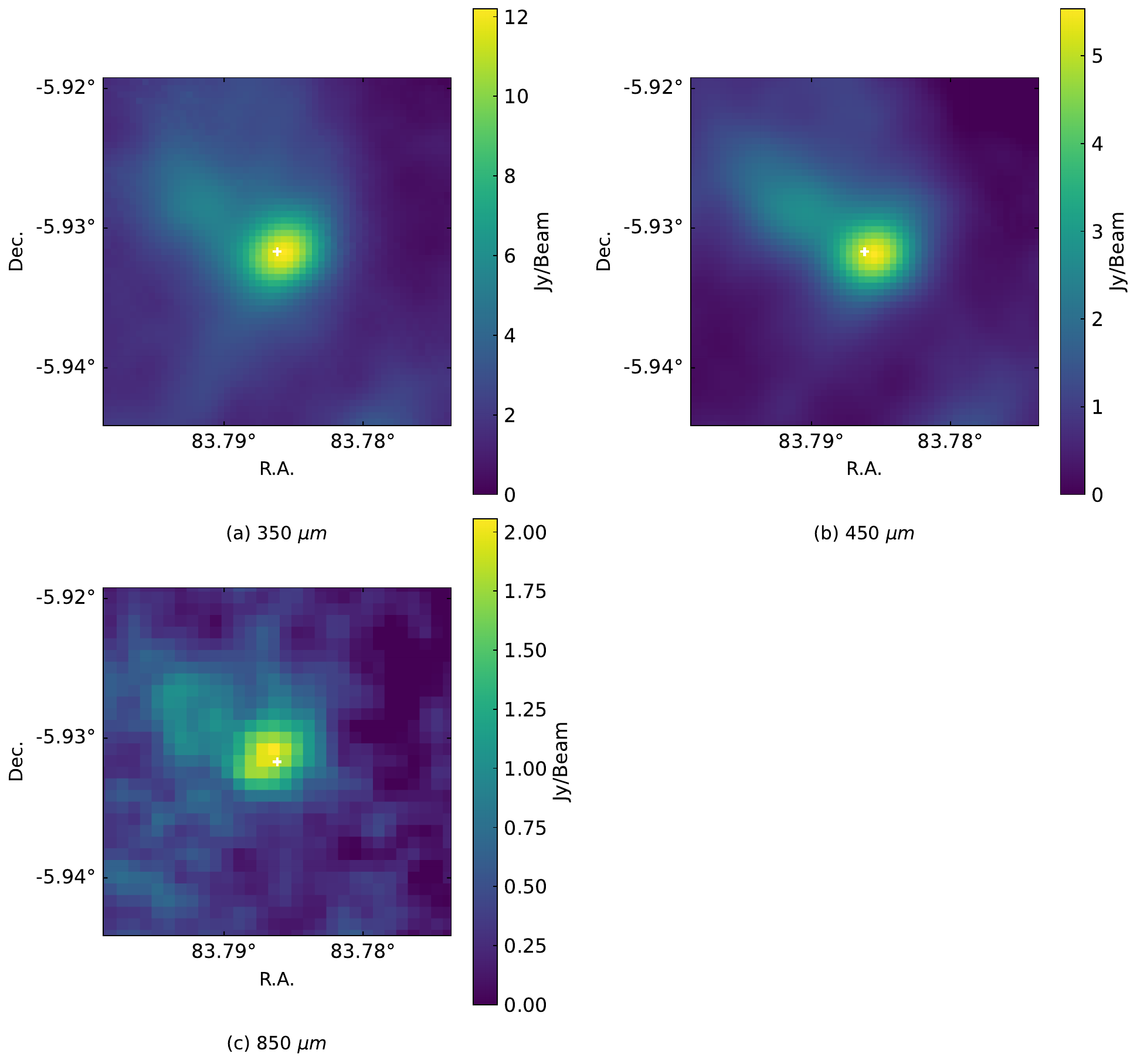}
  \caption{The dust continuum maps of Ori2-2. The maps show the central 0.09 pc region. }
  \label{fig:Ori2-2fig}
\end{figure}

We supplement the three-band continuum images with Spitzer/MIPS 90--97 $\mu$m SED data. This provides data constraints on the Wien side of the SED, helping to break the $\beta$--$T$ degeneracy. CARPP handles such auxiliary SED data by automatically integrating the model images over the corresponding aperture and adding a scaled SED $\chi^2$ term to the total objective function:
\begin{equation}
  \chi^2_{\rm total} = \chi^2_{\rm spatial} + \frac{1}{M_{\rm SED}} \sum_{j=1}^{M_{\rm SED}} \left( \frac{F_{{\rm obs},j} - F_{{\rm model},j}}{\sigma_{{\rm SED},j}} \right)^2,
\end{equation}
where $\chi^2_{\rm spatial}$ is defined in Section~\ref{sec:CARPP_workflow} and $M_{\rm SED}$ is the number of SED points. The factor $1/M_{\rm SED}$ balances the contribution of the few SED points against the many image pixels. This joint fitting effectively breaks the $\beta$--$T$ degeneracy, as shown by the comparison between the with-SED and without-SED results in Table~\ref{tab:real}.

We conducted CARPP fitting for Ori2-2 using 350, 450, and 850 $\mu$m dust continuum maps from CSO and JCMT (Figure~\ref{fig:Ori2-2fig}). The beam sizes of these maps are 8.5$''$, 8.5$''$, and 14.0$''$, respectively. And the measured map RMS noise levels are 0.80, 0.33, and 0.235~Jy/beam. These values correspond to peak signal-to-noise ratios of approximately 15, 17, and 9. To break the $T$--$\beta$ degeneracy, we incorporated additional SED data (52--97~$\mu$m) from the Spitzer/MIPS instrument (PID 30645). CARPP's forward-fitting architecture seamlessly pairs incomplete localized SED data (such as uneven Spitzer patches) with cohesive continuum maps. This maximizes the utilization of precious archival observastions. We compared two models: one excluding and one including the Spitzer SED data (Table~\ref{tab:real}). The core radius, $R_{\rm out}$ = 0.15 pc, was determined from the 850 $\mu$m intensity map. Background and RMS noise levels were measured in regions beyond $R_{\rm out}$. Due to limited map completeness, only data within 0.09 pc from the center were used for fitting. 

The best-fit grid models are [$\beta$, $\alpha$, $r_0$, $T_0$, $T_1$, $r_{\rm t}$]=[1.6, 2.2, 6.25, 30.0, 10.0, 12.5] (without extra SED) and [1.6, 2.4, 3.125, 20.0, 25.0, 12.5] (with extra SED), respectively. The final results are summarized in Table~\ref{tab:real}. As shown in Figure~\ref{fig:nffig}, the most notable difference between the two fittings lies in the temperature profile. The result without extra SED suggests a warmer core with a central temperature of $T_0 = 35.60$ K, while the SED-inclusive result favors a cooler core ($T_0 = 22.79$ K). In both results, the ambient temperature is lower than the central temperature. The differences between the two models in the spectral index and density profiles are relatively minor, with both models exhibiting a flat $\beta$ and nearly identical density profiles in the outer regions. The primary difference is observed in the central density $\rho_0$. In both cases, the density profile is more like a power-law, with a small $r_0$, and spans nearly two orders of magnitude from the center to the edge, surpassing the critical density contrast of the B-E sphere (which is $\rho_{\rm 0}/\rho_R \approx 14.1$). This instability suggests that the density profile of Ori2-2 is more similar to free-fall than hydrostatic equilibrium.

\section{Discussion}\label{sec:discuss}

CARPP provides an efficient and robust approach for deriving the density and temperature distributions of molecular cores from multi-wavelength dust continuum data. By combining a spherically symmetric model with parameterized radiative transfer calculations, CARPP overcomes several key limitations of commonly used SED fitting methods, particularly their inability to handle temperature gradients and optical depth effects along the line of sight. Results from both synthetic and observational tests demonstrate CARPP's ability to recover accurate physical parameters under certain noise and resolution conditions. 
In particular, our realistic applications to the cores TMC-1C and Ori2-2 show that by employing parameterized radiative transfer coupled with a flexible Plummer-like density structure, CARPP can successfully distinguish between gravitationally stable, hydrostatic configurations (such as critical B-E spheres) and dynamically collapsing, free-falling structures. This provides a direct path for researchers to translate multi-wavelength dust continuum maps into meaningful assessments of a core's dynamical and evolutionary status.

We compared CARPP with three other commonly used SED methods using multiple synthetic cores in Section~\ref{sec:4syn}. The single-wavelength `850~$\mu$m estimate' and the multi-wavelength `350--850~$\mu$m fit' represent the simplest approach of single-temperature component modified blackbody fitting. These methods are highly efficient and easy to implement, making them well suited for optically thin regions with modest temperature variations. The multi-wavelength version reduces the number of prior assumptions by leveraging observations from multiple bands. The most straightforward upgrade to single-temperature fitting is multi-component (usually two- or three-temperature) SED modeling (e.g., \citealt{Pagani2015,Miettinen2020}). These studies often use external constraints, such as molecular line observations, to indirectly assume the dust temperatures of the dense central core and the ambient envelope, approximating the line-of-sight emission as a combination of discrete thermal phases. A further advancement is represented by more complex 2D methods with continuous or finely binned multi-temperature distributions, such as PPMAP \citep{Marsh2015}. These methods decompose emission maps into a finely binned grid of temperature and column density components without assuming absolute geometric constraints. These multi-temperature tools provide fine column density maps for molecular clouds based on the optically thin approximation. They can separate the temperature components in the line of sight, but cannot reconstruct the corresponding temperature structures into 3D space. Direct radial inversion methods like the inverse Abel transform \citep{Roy2014} addresses this limitation. This approach allow the dust density and temperature to vary continuously as a function of physical radius $r$, thereby recovering the spatial profiles at the cost of assuming spherical symmetry (or another well-defined geometric template) suitable for core analysis. At the most rigorous end are full, self-consistent 3D dust radiative transfer codes, such as CRT \citep{Juvela2003} and RADMC-3D \citep{Dullemond2012}. While offering high precision, their computational cost is exceptionally high, and Monte Carlo photon noise makes them less practical for direct multi-parameter optimization loops, unless utilizing pre-calculated grids as in \citet{Scibelli2023}, which still requires extensive upfront computation. CARPP occupies a favorable middle ground on this complexity spectrum of SED modeling methods. Assuming a spherically symmetric structure, CARPP is similar to and a little more complex than the inverse Abel transform method, and is suitable for the analysis to molecular cores. Compared to the inverse Abel transform, which is faster but restricted to optically thin cases, CARPP adopts a parameterized forward-modeling approach that solves the full radiative transfer equation. This comes at a moderate computational cost, typically tens of seconds to two minutes per core on a normal personal computer.

While CARPP offers significant advantages over simplified fitting methods, its performance depends sensitively on data quality. High angular resolution and low noise are particularly important for constraining the central density, which is otherwise prone to large uncertainties. Accurate knowledge of the core's total radius is also essential; if the adopted radius is significantly underestimated or overestimated, the fitted density and temperature parameters can be biased. We therefore recommend estimating the core radius from long-wavelength continuum maps or other tracers prior to running CARPP. Among the seven fitted parameters, the spectral index $\beta$ is generally well constrained and exhibits relatively small fitting errors. Fixing $\beta$ using ancillary information—such as a simple SED analysis or CARPP's own grid-search module—can help reduce degeneracies and improve fit stability. Similarly, the central temperature $T_0$ may be constrained independently using molecular line observations (e.g., NH$_3$ or N$_2$H$^+$). Although when using gas kinetic temperatures to constrain dust temperatures, one must account for the fact that the gas and dust temperatures are not necessarily equal \citep{Goldsmith2001, Galli2002}. External constraints like this can help enhancing the robustness of the derived density and temperature profiles, particularly for faint or poorly resolved cores.

Moreover, to reduce the number of free parameters and optimize computational efficiency, several model constants (such as the dust absorption efficiency coefficient \(Q_{350}\) or the opacity \(\kappa_{350}\), the default grain radius \(r_{\rm d}\), the gas-to-dust mass ratio \(R_{\rm g-d}\), and the temperature profile index \(\alpha_{\rm t}\)) are kept fixed in the CARPP fitting process. In our synthetic test cases, these constants are assigned their exact values as defined in the input synthetic profiles, which are generally unknown in actual observations. For real-world observations, these parameters carry significant physical uncertainties. For instance, the submillimeter dust absorption coefficient/opacity alone is estimated to have a regional uncertainty of typically up to a factor of \(\sim2\) \citep{Ossenkopf1994}, and can reach a factor of 8–10 in extreme cases (see the \(Q_{350}\) values listed in Section~\ref{sec:21}). Since the derived density scale (e.g., the central density \(\rho_0\)) is tied to these scaling factors, this uncertainty leads to a systematic error of similar magnitude on the densities. In optically thin regimes, the error is directly proportional to the scaling factors. These issues are common to all SED-fitting methods and are inherently difficult to resolve. We recommend that users carefully select the fixed constant parameters based on regional cloud conditions, or employ grid-search optimization to explore the parameter space of these constants if needed.

Several extensions could further improve the capabilities and flexibility of the CARPP framework. The set of analytical fitting functions can be expanded beyond the current profiles. Using more complex forms inspired by numerical models, such as Larson-Penston collapse density fields \citep{Larson1969,Penston1969} or custom profile templates from hydrodynamic simulations of core formation, would let users test specific dynamical scenarios directly. Also, while the assumption of spherical symmetry is a useful first‑order approximation for many isolated cores, prestellar and protostellar cores can have non‑spherical shapes (e.g., triaxial or spheroidal). Extending CARPP to such geometries, like oblate or prolate spheroids, would therefore be valuable. This could be done by adding coordinate transformation parameters (e.g., inclination angle and aspect ratio) to the line‑of‑sight integration, allowing 3D spheroidal structure reconstruction without losing the computational efficiency of the forward parameterized model.

\section{Conclusion}\label{sec:summary}

CARPP is an open-source, efficient core fitting package designed to recover the density and temperature profiles of molecular cores from multi-wavelength dust continuum observations. By solving the 1D layered radiative transfer equation, CARPP addresses key physical limitations of simple 2D SED fits, which fail to account for line-of-sight temperature gradients and optical depth effects. The main achievements and features of the CARPP framework are:
\begin{enumerate}
\item CARPP is highly robust in parameter recovery, achieving average relative parameter errors of $<20\%$ under typical noise and resolution conditions. Crucially, by fitting all wavelengths in a unified model, CARPP leverages high-spatial-resolution short-wavelength observations to compensate for poorer resolution at longer wavelengths, helping the reconstruction of central core properties.
\item As a forward‑modeling framework, CARPP naturally accommodates heterogeneous datasets with different pixel grids, fields of view, and resolutions. It can directly incorporate localized spectral measurements (e.g., Spitzer/MIPS SED data) without regridding or down‑sampling. The Powell multidimensional minimization algorithm delivers fast convergence within tens of seconds. Moreover, CARPP natively provides reduced $\chi^2$ values, enabling a quantitative assessment of the reliability of the reconstructed density and temperature profiles.
\item CARPP employs generalized Plummer-like density profiles and continuous temperature functions, enabling it to model both hydrostatic equilibrium configurations with B-E profiles and collapsing cores with power-law-dominated profiles. Its high computational efficiency makes it suitable for processing large amounts of data, enabling efficient identification of core density, temperature profiles, and derive the corresponding evolutionary stages. 
\item Applying CARPP to TMC-1C and Ori2-2 successfully demonstrates its ability to distinguish core dynamical states. TMC-1C is identified as a stable, near-critical Bonnor‑Ebert sphere in hydrostatic equilibrium, while Ori2-2 exhibits a power‑law‑dominated, unstable density profile indicative of free‑fall collapse. 
\end{enumerate}
By fitting multi-wavelength dust continuum data with parameterized radiative transfer, CARPP provides an efficient framework for obtaining molecular cores' density and temperature profiles, and therefore understanding their dynamical states. 

\section*{Acknowledgements}
This work is supported by National Natural Science Foundation of China (NSFC) grants Nos. 12588202. We acknowledge the pioneering contributions of K. Marsh and collaborators to the early development of parametric core fitting methods. The CARPP code shares a common ancestry with the COREFIT algorithm, reflecting collaborative efforts involving K. Marsh, D. Li, and others. We thank them for laying the groundwork upon which the community continues to build. 

\section*{Data Availability}
For TMC-1C, the JCMT SCUBA and IRAM 30-m MAMBO data were obtained from \citet{Schnee2007_0}. The CSO 2100 $\mu$m Bolocam observations of TMC-1C were taken as part of a PI-led programme at CSO (PI: Di Li) and are available from the corresponding author upon reasonable request. For Ori2-2, the 350 and 450 $\mu$m CSO continuum data were taken from \citet{Li2007}, and the 850 $\mu$m data were retrieved from the public JCMT archive. The Spitzer MIPS SED observations were obtained under programme PID 30645 (PI: Di Li) and are accessible through the Spitzer Heritage Archive. The CARPP package developed for this work is publicly available at \url{https://github.com/olozhika/CARPP}. 

\bibliographystyle{mnras}
\bibliography{sample631} 

@ARTICLE{Mathis1977,
       author = {{Mathis}, J.~S. and {Rumpl}, W. and {Nordsieck}, K.~H.},
        title = "{The size distribution of interstellar grains.}",
      journal = {\apj},
         year = 1977,
        month = oct,
       volume = {217},
        pages = {425-433},
          doi = {10.1086/155591},
       adsurl = {https://ui.adsabs.harvard.edu/abs/1977ApJ...217..425M}
}

@ARTICLE{Hildebrand1983,
       author = {{Hildebrand}, R.~H.},
        title = "{The determination of cloud masses and dust characteristics from submillimetre thermal emission.}",
      journal = {\qjras},
         year = 1983,
        month = sep,
       volume = {24},
        pages = {267-282},
       adsurl = {https://ui.adsabs.harvard.edu/abs/1983QJRAS..24..267H}
}

@ARTICLE{Marsh2014,
       author = {{Marsh}, K.~A. and {Griffin}, M.~J. and {Palmeirim}, P. and {Andr{\'e}}, Ph. and {Kirk}, J. and {Stamatellos}, D. and {Ward-Thompson}, D. and {Roy}, A. and {Bontemps}, S. and {di Francesco}, J. and {Elia}, D. and {Hill}, T. and {K{\"o}nyves}, V. and {Motte}, F. and {Nguyen-Luong}, Q. and {Peretto}, N. and {Pezzuto}, S. and {Rivera-Ingraham}, A. and {Schneider}, N. and {Spinoglio}, L. and {White}, G.},
        title = "{Properties of starless and prestellar cores in Taurus revealed by Herschel: SPIRE/PACS imaging}",
      journal = {\mnras},
         year = 2014,
        month = apr,
       volume = {439},
       number = {4},
        pages = {3683-3693},
          doi = {10.1093/mnras/stu219},
archivePrefix = {arXiv},
       eprint = {1401.7871},
 primaryClass = {astro-ph.GA},
       adsurl = {https://ui.adsabs.harvard.edu/abs/2014MNRAS.439.3683M}
}

@ARTICLE{Planck2011,
       author = {{Planck Collaboration} and {Ade}, P.~A.~R. and {Aghanim}, N. and {Arnaud}, M. and {Ashdown}, M. and {Aumont}, J. and {Baccigalupi}, C. and {Balbi}, A. and {Banday}, A.~J. and {Barreiro}, R.~B. and {Bartlett}, J.~G. and {Battaner}, E. and {Benabed}, K. and {Beno{\^\i}t}, A. and {Bernard}, J. -P. and {Bersanelli}, M. and {Bhatia}, R. and {Bock}, J.~J. and {Bonaldi}, A. and {Bond}, J.~R. and {Borrill}, J. and {Bouchet}, F.~R. and {Boulanger}, F. and {Bucher}, M. and {Burigana}, C. and {Cabella}, P. and {Cantalupo}, C.~M. and {Cardoso}, J. -F. and {Catalano}, A. and {Cay{\'o}n}, L. and {Challinor}, A. and {Chamballu}, A. and {Chiang}, L. -Y. and {Christensen}, P.~R. and {Clements}, D.~L. and {Colombi}, S. and {Couchot}, F. and {Coulais}, A. and {Crill}, B.~P. and {Cuttaia}, F. and {Danese}, L. and {Davies}, R.~D. and {de Bernardis}, P. and {de Gasperis}, G. and {de Rosa}, A. and {de Zotti}, G. and {Delabrouille}, J. and {Delouis}, J. -M. and {D{\'e}sert}, F. -X. and {Dickinson}, C. and {Doi}, Y. and {Donzelli}, S. and {Dor{\'e}}, O. and {D{\"o}rl}, U. and {Douspis}, M. and {Dupac}, X. and {Efstathiou}, G. and {En{\ss}lin}, T.~A. and {Falgarone}, E. and {Finelli}, F. and {Forni}, O. and {Frailis}, M. and {Franceschi}, E. and {Galeotta}, S. and {Ganga}, K. and {Giard}, M. and {Giardino}, G. and {Giraud-H{\'e}raud}, Y. and {Gonz{\'a}lez-Nuevo}, J. and {G{\'o}rski}, K.~M. and {Gratton}, S. and {Gregorio}, A. and {Gruppuso}, A. and {Hansen}, F.~K. and {Harrison}, D. and {Helou}, G. and {Henrot-Versill{\'e}}, S. and {Herranz}, D. and {Hildebrandt}, S.~R. and {Hivon}, E. and {Hobson}, M. and {Holmes}, W.~A. and {Hovest}, W. and {Hoyland}, R.~J. and {Huffenberger}, K.~M. and {Ikeda}, N. and {Jaffe}, A.~H. and {Jones}, W.~C. and {Juvela}, M. and {Keih{\"a}nen}, E. and {Keskitalo}, R. and {Kisner}, T.~S. and {Kitamura}, Y. and {Kneissl}, R. and {Knox}, L. and {Kurki-Suonio}, H. and {Lagache}, G. and {Lamarre}, J. -M. and {Lasenby}, A. and {Laureijs}, R.~J. and {Lawrence}, C.~R. and {Leach}, S. and {Leonardi}, R. and {Leroy}, C. and {Linden-V{\o}rnle}, M. and {L{\'o}pez-Caniego}, M. and {Lubin}, P.~M. and {Mac{\'\i}as-P{\'e}rez}, J.~F. and {MacTavish}, C.~J. and {Maffei}, B. and {Malinen}, J. and {Mandolesi}, N. and {Mann}, R. and {Maris}, M. and {Marshall}, D.~J. and {Martin}, P. and {Mart{\'\i}nez-Gonz{\'a}lez}, E. and {Masi}, S. and {Matarrese}, S. and {Matthai}, F. and {Mazzotta}, P. and {McGehee}, P. and {Melchiorri}, A. and {Mendes}, L. and {Mennella}, A. and {Meny}, C. and {Mitra}, S. and {Miville-Desch{\^e}nes}, M. -A. and {Moneti}, A. and {Montier}, L. and {Morgante}, G. and {Mortlock}, D. and {Munshi}, D. and {Murphy}, A. and {Naselsky}, P. and {Nati}, F. and {Natoli}, P. and {Netterfield}, C.~B. and {N{\o}rgaard-Nielsen}, H.~U. and {Noviello}, F. and {Novikov}, D. and {Novikov}, I. and {Osborne}, S. and {Pagani}, L. and {Pajot}, F. and {Paladini}, R. and {Pasian}, F. and {Patanchon}, G. and {Pelkonen}, V. -M. and {Perdereau}, O. and {Perotto}, L. and {Perrotta}, F. and {Piacentini}, F. and {Piat}, M. and {Plaszczynski}, S. and {Pointecouteau}, E. and {Polenta}, G. and {Ponthieu}, N. and {Poutanen}, T. and {Pr{\'e}zeau}, G. and {Prunet}, S. and {Puget}, J. -L. and {Reach}, W.~T. and {Rebolo}, R. and {Reinecke}, M. and {Renault}, C. and {Ricciardi}, S. and {Riller}, T. and {Ristorcelli}, I. and {Rocha}, G. and {Rosset}, C. and {Rowan-Robinson}, M. and {Rubi{\~n}o-Mart{\'\i}n}, J.~A. and {Rusholme}, B. and {Sandri}, M. and {Santos}, D. and {Savini}, G. and {Scott}, D. and {Seiffert}, M.~D. and {Smoot}, G.~F. and {Starck}, J. -L. and {Stivoli}, F. and {Stolyarov}, V. and {Sudiwala}, R. and {Sygnet}, J. -F. and {Tauber}, J.~A. and {Terenzi}, L. and {Toffolatti}, L. and {Tomasi}, M. and {Torre}, J. -P. and {Toth}, V. and {Tristram}, M. and {Tuovinen}, J. and {Umana}, G. and {Valenziano}, L. and {Vielva}, P. and {Villa}, F. and {Vittorio}, N. and {Wade}, L.~A. and {Wandelt}, B.~D.},
        title = "{Planck early results. XXII. The submillimetre properties of a sample of Galactic cold clumps}",
      journal = {\aap},
         year = 2011,
        month = dec,
       volume = {536},
          eid = {A22},
        pages = {A22},
          doi = {10.1051/0004-6361/201116481},
archivePrefix = {arXiv},
       eprint = {1101.2034},
 primaryClass = {astro-ph.GA},
       adsurl = {https://ui.adsabs.harvard.edu/abs/2011A&A...536A..22P}
}

@ARTICLE{Kelly2012,
       author = {{Kelly}, Brandon C. and {Shetty}, Rahul and {Stutz}, Amelia M. and {Kauffmann}, Jens and {Goodman}, Alyssa A. and {Launhardt}, Ralf},
        title = "{Dust Spectral Energy Distributions in the Era of Herschel and Planck: A Hierarchical Bayesian-fitting Technique}",
      journal = {\apj},
         year = 2012,
        month = jun,
       volume = {752},
       number = {1},
          eid = {55},
        pages = {55},
          doi = {10.1088/0004-637X/752/1/55},
archivePrefix = {arXiv},
       eprint = {1203.0025},
 primaryClass = {astro-ph.IM},
       adsurl = {https://ui.adsabs.harvard.edu/abs/2012ApJ...752...55K}
}

@ARTICLE{Marsh2015,
       author = {{Marsh}, K.~A. and {Whitworth}, A.~P. and {Lomax}, O.},
        title = "{Temperature as a third dimension in column-density mapping of dusty astrophysical structures associated with star formation}",
      journal = {\mnras},
         year = 2015,
        month = dec,
       volume = {454},
       number = {4},
        pages = {4282-4292},
          doi = {10.1093/mnras/stv2248},
archivePrefix = {arXiv},
       eprint = {1509.08699},
 primaryClass = {astro-ph.IM},
       adsurl = {https://ui.adsabs.harvard.edu/abs/2015MNRAS.454.4282M}
}

@ARTICLE{Goldsmith1997,
       author = {{Goldsmith}, Paul F. and {Bergin}, Edwin A. and {Lis}, D.~C.},
        title = "{Carbon Monoxide and Dust Column Densities: The Dust-to-Gas Ratio and Structure of Three Giant Molecular Cloud Cores}",
      journal = {\apj},
         year = 1997,
        month = dec,
       volume = {491},
       number = {2},
        pages = {615-637},
          doi = {10.1086/304986},
       adsurl = {https://ui.adsabs.harvard.edu/abs/1997ApJ...491..615G}
}

@ARTICLE{Draine1984,
       author = {{Draine}, B.~T. and {Lee}, H.~M.},
        title = "{Optical Properties of Interstellar Graphite and Silicate Grains}",
      journal = {\apj},
         year = 1984,
        month = oct,
       volume = {285},
        pages = {89},
          doi = {10.1086/162480},
       adsurl = {https://ui.adsabs.harvard.edu/abs/1984ApJ...285...89D}
}

@ARTICLE{Pollack1994,
       author = {{Pollack}, James B. and {Hollenbach}, David and {Beckwith}, Steven and {Simonelli}, Damon P. and {Roush}, Ted and {Fong}, Wesley},
        title = "{Composition and Radiative Properties of Grains in Molecular Clouds and Accretion Disks}",
      journal = {\apj},
         year = 1994,
        month = feb,
       volume = {421},
        pages = {615},
          doi = {10.1086/173677},
       adsurl = {https://ui.adsabs.harvard.edu/abs/1994ApJ...421..615P}
}

@ARTICLE{Montillaud2015,
       author = {{Montillaud}, J. and {Juvela}, M. and {Rivera-Ingraham}, A. and {Malinen}, J. and {Pelkonen}, V. -M. and {Ristorcelli}, I. and {Montier}, L. and {Marshall}, D.~J. and {Marton}, G. and {Pagani}, L. and {Toth}, L.~V. and {Zahorecz}, S. and {Ysard}, N. and {McGehee}, P. and {Paladini}, R. and {Falgarone}, E. and {Bernard}, J. -P. and {Motte}, F. and {Zavagno}, A. and {Doi}, Y.},
        title = "{Galactic cold cores. IV. Cold submillimetre sources: catalogue and statistical analysis}",
      journal = {\aap},
         year = 2015,
        month = dec,
       volume = {584},
          eid = {A92},
        pages = {A92},
          doi = {10.1051/0004-6361/201424063},
       adsurl = {https://ui.adsabs.harvard.edu/abs/2015A&A...584A..92M}
}

@ARTICLE{Preibisch1993,
       author = {{Preibisch}, Th. and {Ossenkopf}, V. and {Yorke}, H.~W. and {Henning}, Th.},
        title = "{The influence of ice-coated grains on protostellar spectra.}",
      journal = {\aap},
         year = 1993,
        month = nov,
       volume = {279},
        pages = {577-588},
       adsurl = {https://ui.adsabs.harvard.edu/abs/1993A&A...279..577P}
}

@ARTICLE{Ossenkopf1994,
       author = {{Ossenkopf}, V. and {Henning}, Th.},
        title = "{Dust opacities for protostellar cores.}",
      journal = {\aap},
         year = 1994,
        month = nov,
       volume = {291},
        pages = {943-959},
       adsurl = {https://ui.adsabs.harvard.edu/abs/1994A&A...291..943O}
}

@ARTICLE{Alves2001,
       author = {{Alves}, Jo{\~a}o F. and {Lada}, Charles J. and {Lada}, Elizabeth A.},
        title = "{Internal structure of a cold dark molecular cloud inferred from the extinction of background starlight}",
      journal = {\nat},
         year = 2001,
        month = jan,
       volume = {409},
       number = {6817},
        pages = {159-161},
          doi = {10.1038/35051509},
       adsurl = {https://ui.adsabs.harvard.edu/abs/2001Natur.409..159A}
}

@ARTICLE{Boudet2005,
       author = {{Boudet}, N. and {Mutschke}, H. and {Nayral}, C. and {J{\"a}ger}, C. and {Bernard}, J.-P. and {Henning}, T. and {Meny}, C.},
        title = "{Temperature Dependence of the Submillimeter Absorption Coefficient of Amorphous Silicate Grains}",
      journal = {\apj},
         year = 2005,
        month = nov,
       volume = {633},
       number = {1},
        pages = {272-281},
          doi = {10.1086/432966},
       adsurl = {https://ui.adsabs.harvard.edu/abs/2005ApJ...633..272B}
}

@ARTICLE{Pattle2015,
       author = {{Pattle}, K. and {Ward-Thompson}, D. and {Kirk}, J.~M. and {White}, G.~J. and {Drabek-Maunder}, E. and {Buckle}, J. and {Beaulieu}, S.~F. and {Berry}, D.~S. and {Broekhoven-Fiene}, H. and {Currie}, M.~J. and {Fich}, M. and {Hatchell}, J. and {Kirk}, H. and {Jenness}, T. and {Johnstone}, D. and {Mottram}, J.~C. and {Nutter}, D. and {Pineda}, J.~E. and {Quinn}, C. and {Salji}, C. and {Tisi}, S. and {Walker-Smith}, S. and {di Francesco}, J. and {Hogerheijde}, M.~R. and {Andr{\'e}}, Ph. and {Bastien}, P. and {Bresnahan}, D. and {Butner}, H. and {Chen}, M. and {Chrysostomou}, A. and {Coude}, S. and {Davis}, C.~J. and {Duarte-Cabral}, A. and {Fiege}, J. and {Friberg}, P. and {Friesen}, R. and {Fuller}, G.~A. and {Graves}, S. and {Greaves}, J. and {Gregson}, J. and {Griffin}, M.~J. and {Holland}, W. and {Joncas}, G. and {Knee}, L.~B.~G. and {K{\"o}nyves}, V. and {Mairs}, S. and {Marsh}, K. and {Matthews}, B.~C. and {Moriarty-Schieven}, G. and {Rawlings}, J. and {Richer}, J. and {Robertson}, D. and {Rosolowsky}, E. and {Rumble}, D. and {Sadavoy}, S. and {Spinoglio}, L. and {Thomas}, H. and {Tothill}, N. and {Viti}, S. and {Wouterloot}, J. and {Yates}, J. and {Zhu}, M.},
        title = "{The JCMT Gould Belt Survey: first results from the SCUBA-2 observations of the Ophiuchus molecular cloud and a virial analysis of its prestellar core population}",
      journal = {\mnras},
         year = 2015,
        month = jun,
       volume = {450},
       number = {1},
        pages = {1094-1122},
          doi = {10.1093/mnras/stv376},
archivePrefix = {arXiv},
       eprint = {1502.05858},
 primaryClass = {astro-ph.GA},
       adsurl = {https://ui.adsabs.harvard.edu/abs/2015MNRAS.450.1094P}
}

@ARTICLE{Kirk2016,
       author = {{Kirk}, H. and {Johnstone}, D. and {Di Francesco}, J. and {Lane}, J. and {Buckle}, J. and {Berry}, D.~S. and {Broekhoven-Fiene}, H. and {Currie}, M.~J. and {Fich}, M. and {Hatchell}, J. and {Jenness}, T. and {Mottram}, J.~C. and {Nutter}, D. and {Pattle}, K. and {Pineda}, J.~E. and {Quinn}, C. and {Salji}, C. and {Tisi}, S. and {Hogerheijde}, M.~R. and {Ward-Thompson}, D. and {JCMT Gould Belt Survey Team}},
        title = "{The JCMT Gould Belt Survey: Dense Core Clusters in Orion B}",
      journal = {\apj},
         year = 2016,
        month = apr,
       volume = {821},
       number = {2},
          eid = {98},
        pages = {98},
          doi = {10.3847/0004-637X/821/2/98},
archivePrefix = {arXiv},
       eprint = {1602.00707},
 primaryClass = {astro-ph.SR},
       adsurl = {https://ui.adsabs.harvard.edu/abs/2016ApJ...821...98K}
}

@ARTICLE{Ward-Thompson2007,
       author = {{Ward-Thompson}, D. and {Di Francesco}, J. and {Hatchell}, J. and {Hogerheijde}, M.~R. and {Nutter}, D. and {Bastien}, P. and {Basu}, Shantanu and {Bonnell}, I. and {Bowey}, J. and {Brunt}, C. and {Buckle}, J. and {Butner}, H. and {Cavanagh}, B. and {Chrysostomou}, A. and {Curtis}, E. and {Davis}, C.~J. and {Dent}, W.~R.~F. and {van Dishoeck}, E. and {Edmunds}, M.~G. and {Fich}, M. and {Fiege}, J. and {Fissel}, L. and {Friberg}, P. and {Friesen}, R. and {Frieswijk}, W. and {Fuller}, G.~A. and {Gosling}, A. and {Graves}, S. and {Greaves}, J.~S. and {Helmich}, F. and {Hills}, R.~E. and {Holland}, W.~S. and {Houde}, M. and {Jayawardhana}, R. and {Johnstone}, D. and {Joncas}, G. and {Kirk}, H. and {Kirk}, J.~M. and {Knee}, L.~B.~G. and {Matthews}, B. and {Matthews}, H. and {Matzner}, C. and {Moriarty-Schieven}, G.~H. and {Naylor}, D. and {Padman}, R. and {Plume}, R. and {Rawlings}, J.~M.~C. and {Redman}, R.~O. and {Reid}, M. and {Richer}, J.~S. and {Shipman}, R. and {Simpson}, R.~J. and {Spaans}, M. and {Stamatellos}, D. and {Tsamis}, Y.~G. and {Viti}, S. and {Weferling}, B. and {White}, G.~J. and {Whitworth}, A.~P. and {Wouterloot}, J. and {Yates}, J. and {Zhu}, M.},
        title = "{The James Clerk Maxwell Telescope Legacy Survey of Nearby Star-forming Regions in the Gould Belt}",
      journal = {\pasp},
         year = 2007,
        month = aug,
       volume = {119},
       number = {858},
        pages = {855-870},
          doi = {10.1086/521277},
archivePrefix = {arXiv},
       eprint = {0707.0169},
 primaryClass = {astro-ph},
       adsurl = {https://ui.adsabs.harvard.edu/abs/2007PASP..119..855W}
}

@ARTICLE{Andre2010,
       author = {{Andr{\'e}}, Ph. and {Men'shchikov}, A. and {Bontemps}, S. and {K{\"o}nyves}, V. and {Motte}, F. and {Schneider}, N. and {Didelon}, P. and {Minier}, V. and {Saraceno}, P. and {Ward-Thompson}, D. and {di Francesco}, J. and {White}, G. and {Molinari}, S. and {Testi}, L. and {Abergel}, A. and {Griffin}, M. and {Henning}, Th. and {Royer}, P. and {Mer{\'\i}n}, B. and {Vavrek}, R. and {Attard}, M. and {Arzoumanian}, D. and {Wilson}, C.~D. and {Ade}, P. and {Aussel}, H. and {Baluteau}, J.-P. and {Benedettini}, M. and {Bernard}, J.-Ph. and {Blommaert}, J.~A.~D.~L. and {Cambr{\'e}sy}, L. and {Cox}, P. and {di Giorgio}, A. and {Hargrave}, P. and {Hennemann}, M. and {Huang}, M. and {Kirk}, J. and {Krause}, O. and {Launhardt}, R. and {Leeks}, S. and {Le Pennec}, J. and {Li}, J.~Z. and {Martin}, P.~G. and {Maury}, A. and {Olofsson}, G. and {Omont}, A. and {Peretto}, N. and {Pezzuto}, S. and {Prusti}, T. and {Roussel}, H. and {Russeil}, D. and {Sauvage}, M. and {Sibthorpe}, B. and {Sicilia-Aguilar}, A. and {Spinoglio}, L. and {Waelkens}, C. and {Woodcraft}, A. and {Zavagno}, A.},
        title = "{From filamentary clouds to prestellar cores to the stellar IMF: Initial highlights from the Herschel Gould Belt Survey}",
      journal = {\aap},
         year = 2010,
        month = jul,
       volume = {518},
          eid = {L102},
        pages = {L102},
          doi = {10.1051/0004-6361/201014666},
archivePrefix = {arXiv},
       eprint = {1005.2618},
 primaryClass = {astro-ph.GA},
       adsurl = {https://ui.adsabs.harvard.edu/abs/2010A&A...518L.102A}
}

@ARTICLE{Ysard2019,
       author = {{Ysard}, N. and {Koehler}, M. and {Jimenez-Serra}, I. and {Jones}, A.~P. and {Verstraete}, L.},
        title = "{From grains to pebbles: the influence of size distribution and chemical composition on dust emission properties}",
      journal = {\aap},
         year = 2019,
        month = nov,
       volume = {631},
          eid = {A88},
        pages = {A88},
          doi = {10.1051/0004-6361/201936089},
archivePrefix = {arXiv},
       eprint = {1909.05015},
 primaryClass = {astro-ph.GA},
       adsurl = {https://ui.adsabs.harvard.edu/abs/2019A&A...631A..88Y}
}

@ARTICLE{Meny2007,
       author = {{Meny}, C. and {Gromov}, V. and {Boudet}, N. and {Bernard}, J.-Ph. and {Paradis}, D. and {Nayral}, C.},
        title = "{Far-infrared to millimeter astrophysical dust emission. I. A model based on physical properties of amorphous solids}",
      journal = {\aap},
         year = 2007,
        month = jun,
       volume = {468},
       number = {1},
        pages = {171-188},
          doi = {10.1051/0004-6361:20065771},
archivePrefix = {arXiv},
       eprint = {astro-ph/0701226},
 primaryClass = {astro-ph},
       adsurl = {https://ui.adsabs.harvard.edu/abs/2007A&A...468..171M}
}

@ARTICLE{Juvela2003,
       author = {{Juvela}, M. and {Padoan}, P.},
        title = "{Dust emission from inhomogeneous interstellar clouds: Radiative transfer in 3D with transiently heated particles}",
      journal = {\aap},
         year = 2003,
        month = jan,
       volume = {397},
        pages = {201-212},
          doi = {10.1051/0004-6361:20021433},
archivePrefix = {arXiv},
       eprint = {astro-ph/0207379},
 primaryClass = {astro-ph},
       adsurl = {https://ui.adsabs.harvard.edu/abs/2003A&A...397..201J}
}

@ARTICLE{Goldsmith2001,
       author = {{Goldsmith}, Paul F.},
        title = "{Molecular Depletion and Thermal Balance in Dark Cloud Cores}",
      journal = {\apj},
         year = 2001,
        month = aug,
       volume = {557},
       number = {2},
        pages = {736-746},
          doi = {10.1086/322255},
       adsurl = {https://ui.adsabs.harvard.edu/abs/2001ApJ...557..736G}
}

@ARTICLE{Galli2002,
       author = {{Galli}, D. and {Walmsley}, M. and {Gon{\c{c}}alves}, J.},
        title = "{The structure and stability of molecular cloud cores in external radiation fields}",
      journal = {\aap},
         year = 2002,
        month = oct,
       volume = {394},
        pages = {275-284},
          doi = {10.1051/0004-6361:20021125},
archivePrefix = {arXiv},
       eprint = {astro-ph/0208416},
 primaryClass = {astro-ph},
       adsurl = {https://ui.adsabs.harvard.edu/abs/2002A&A...394..275G}
}

@article{Pirogov2009,
archivePrefix = {arXiv},
arxivId = {0911.4421},
author = {Pirogov, L. E.},
doi = {10.1134/S1063772909120051},
eprint = {0911.4421},
issn = {10637729},
journal = {Astronomy Reports},
number = {12},
pages = {1127--1135},
title = {{Density profiles in molecular cloud cores associated with high-mass star-forming regions}},
volume = {53},
year = {2009}
}

@ARTICLE{McCrea1957,
       author = {{McCrea}, W.~H.},
        title = "{The formation of Population I Stars. Part I. Gravitational contraction}",
      journal = {\mnras},
         year = 1957,
        month = jan,
       volume = {117},
        pages = {562},
          doi = {10.1093/mnras/117.5.562},
       adsurl = {https://ui.adsabs.harvard.edu/abs/1957MNRAS.117..562M}
}

@article{Tafalla2004,
archivePrefix = {arXiv},
arxivId = {astro-ph/0401148},
author = {Tafalla, M. and Myers, P. C. and Caselli, P. and Walmsley, C. M.},
doi = {10.1023/B:ASTR.0000045036.76044.bd},
eprint = {0401148},
issn = {0004640X},
journal = {Astrophysics and Space Science},
number = {1-4},
pages = {347--354},
primaryClass = {astro-ph},
title = {{On the internal structure of starless cores. Physical and chemical properties of L1498 and L1517B}},
volume = {292},
year = {2004}
}

@ARTICLE{Kirk2005,
       author = {{Kirk}, J.~M. and {Ward-Thompson}, D. and {Andr{\'e}}, P.},
        title = "{The initial conditions of isolated star formation - VI. SCUBA mappingof pre-stellar cores}",
      journal = {\mnras},
         year = 2005,
        month = jul,
       volume = {360},
       number = {4},
        pages = {1506-1526},
          doi = {10.1111/j.1365-2966.2005.09145.x},
archivePrefix = {arXiv},
       eprint = {astro-ph/0505190},
 primaryClass = {astro-ph},
       adsurl = {https://ui.adsabs.harvard.edu/abs/2005MNRAS.360.1506K}
}

@ARTICLE{Pagani2015,
       author = {{Pagani}, L. and {Lef{\`e}vre}, C. and {Juvela}, M. and {Pelkonen}, V. -M. and {Schuller}, F.},
        title = "{Can we trace very cold dust from its emission alone?}",
      journal = {\aap},
         year = 2015,
        month = feb,
       volume = {574},
          eid = {L5},
        pages = {L5},
          doi = {10.1051/0004-6361/201425095},
archivePrefix = {arXiv},
       eprint = {1501.00861},
 primaryClass = {astro-ph.GA},
       adsurl = {https://ui.adsabs.harvard.edu/abs/2015A&A...574L...5P}
}

@ARTICLE{Bonnor1956,
       author = {{Bonnor}, W.~B.},
        title = "{Boyle's Law and gravitational instability}",
      journal = {\mnras},
         year = 1956,
        month = jan,
       volume = {116},
        pages = {351},
          doi = {10.1093/mnras/116.3.351},
       adsurl = {https://ui.adsabs.harvard.edu/abs/1956MNRAS.116..351B}
}

@ARTICLE{Larson1972,
       author = {{Larson}, Richard B.},
        title = "{The Collapse of a Rotating Cloud}",
      journal = {\mnras},
         year = 1972,
        month = jan,
       volume = {156},
        pages = {437},
          doi = {10.1093/mnras/156.4.437},
       adsurl = {https://ui.adsabs.harvard.edu/abs/1972MNRAS.156..437L}
}

@INPROCEEDINGS{Krumholz2005,
       author = {{Krumholz}, Mark R. and {Klein}, Richard I. and {McKee}, Christopher F.},
        title = "{Radiation pressure in massive star formation}",
    booktitle = {Massive Star Birth: A Crossroads of Astrophysics},
         year = 2005,
       editor = {{Cesaroni}, R. and {Felli}, M. and {Churchwell}, E. and {Walmsley}, M.},
       series = {IAU Symposium},
       volume = {227},
        month = jan,
        pages = {231-236},
          doi = {10.1017/S1743921305004588},
archivePrefix = {arXiv},
       eprint = {astro-ph/0510432},
 primaryClass = {astro-ph},
       adsurl = {https://ui.adsabs.harvard.edu/abs/2005IAUS..227..231K}
}

@ARTICLE{Bhandare2018,
       author = {{Bhandare}, Asmita and {Kuiper}, Rolf and {Henning}, Thomas and {Fendt}, Christian and {Marleau}, Gabriel-Dominique and {K{\"o}lligan}, Anders},
        title = "{First core properties: from low- to high-mass star formation}",
      journal = {\aap},
         year = 2018,
        month = oct,
       volume = {618},
          eid = {A95},
        pages = {A95},
          doi = {10.1051/0004-6361/201832635},
archivePrefix = {arXiv},
       eprint = {1807.06597},
 primaryClass = {astro-ph.SR},
       adsurl = {https://ui.adsabs.harvard.edu/abs/2018A&A...618A..95B}
}

@ARTICLE{Hung2010,
       author = {{Hung}, Chao-Ling and {Lai}, Shih-Ping and {Yan}, Chi-Hung},
        title = "{The Evolution of Density Structure of Starless and Protostellar Cores}",
      journal = {\apj},
         year = 2010,
        month = feb,
       volume = {710},
       number = {1},
        pages = {207-211},
          doi = {10.1088/0004-637X/710/1/207},
archivePrefix = {arXiv},
       eprint = {0912.4738},
 primaryClass = {astro-ph.SR},
       adsurl = {https://ui.adsabs.harvard.edu/abs/2010ApJ...710..207H}
}

@ARTICLE{Larson1969,
       author = {{Larson}, Richard B.},
        title = "{Numerical calculations of the dynamics of collapsing proto-star}",
      journal = {\mnras},
         year = 1969,
        month = jan,
       volume = {145},
        pages = {271},
          doi = {10.1093/mnras/145.3.271},
       adsurl = {https://ui.adsabs.harvard.edu/abs/1969MNRAS.145..271L}
}

@ARTICLE{Penston1969,
       author = {{Penston}, M.~V.},
        title = "{Dynamics of self-gravitating gaseous spheres-III. Analytical results in the free-fall of isothermal cases}",
      journal = {\mnras},
         year = 1969,
        month = jan,
       volume = {144},
        pages = {425},
          doi = {10.1093/mnras/144.4.425},
       adsurl = {https://ui.adsabs.harvard.edu/abs/1969MNRAS.144..425P}
}

@ARTICLE{Andre1996,
       author = {{Andre}, P. and {Ward-Thompson}, D. and {Motte}, F.},
        title = "{Probing the initial conditions of star formation: the structure of the prestellar core L 1689B.}",
      journal = {\aap},
         year = 1996,
        month = oct,
       volume = {314},
        pages = {625-635},
       adsurl = {https://ui.adsabs.harvard.edu/abs/1996A&A...314..625A}
}

@ARTICLE{Ward-Thompson1994,
       author = {{Ward-Thompson}, D. and {Scott}, P.~F. and {Hills}, R.~E. and {Andre}, P.},
        title = "{A Submillimetre Continuum Survey of Pre Protostellar Cores}",
      journal = {\mnras},
         year = 1994,
        month = may,
       volume = {268},
        pages = {276},
          doi = {10.1093/mnras/268.1.276},
       adsurl = {https://ui.adsabs.harvard.edu/abs/1994MNRAS.268..276W}
}

@ARTICLE{Beuther2002,
       author = {{Beuther}, H. and {Schilke}, P. and {Menten}, K.~M. and {Motte}, F. and {Sridharan}, T.~K. and {Wyrowski}, F.},
        title = "{High-Mass Protostellar Candidates. II. Density Structure from Dust Continuum and CS Emission}",
      journal = {\apj},
         year = 2002,
        month = feb,
       volume = {566},
       number = {2},
        pages = {945-965},
          doi = {10.1086/338334},
archivePrefix = {arXiv},
       eprint = {astro-ph/0110370},
 primaryClass = {astro-ph},
       adsurl = {https://ui.adsabs.harvard.edu/abs/2002ApJ...566..945B}
}

@ARTICLE{Caselli2002,
       author = {{Caselli}, Paola and {Benson}, Priscilla J. and {Myers}, Philip C. and {Tafalla}, Mario},
        title = "{Dense Cores in Dark Clouds. XIV. N$_{2}$H$^{+}$ (1-0) Maps of Dense Cloud Cores}",
      journal = {\apj},
         year = 2002,
        month = jun,
       volume = {572},
       number = {1},
        pages = {238-263},
          doi = {10.1086/340195},
archivePrefix = {arXiv},
       eprint = {astro-ph/0202173},
 primaryClass = {astro-ph},
       adsurl = {https://ui.adsabs.harvard.edu/abs/2002ApJ...572..238C}
}

@ARTICLE{Launhardt2013,
       author = {{Launhardt}, R. and {Stutz}, A.~M. and {Schmiedeke}, A. and {Henning}, Th. and {Krause}, O. and {Balog}, Z. and {Beuther}, H. and {Birkmann}, S. and {Hennemann}, M. and {Kainulainen}, J. and {Khanzadyan}, T. and {Linz}, H. and {Lippok}, N. and {Nielbock}, M. and {Pitann}, J. and {Ragan}, S. and {Risacher}, C. and {Schmalzl}, M. and {Shirley}, Y.~L. and {Stecklum}, B. and {Steinacker}, J. and {Tackenberg}, J.},
        title = "{The Earliest Phases of Star Formation (EPoS): a Herschel key project. The thermal structure of low-mass molecular cloud cores}",
      journal = {\aap},
         year = 2013,
        month = mar,
       volume = {551},
          eid = {A98},
        pages = {A98},
          doi = {10.1051/0004-6361/201220477},
archivePrefix = {arXiv},
       eprint = {1301.1498},
 primaryClass = {astro-ph.SR},
       adsurl = {https://ui.adsabs.harvard.edu/abs/2013A&A...551A..98L}
}

@ARTICLE{Roy2014,
       author = {{Roy}, A. and {Andr{\'e}}, Ph. and {Palmeirim}, P. and {Attard}, M. and {K{\"o}nyves}, V. and {Schneider}, N. and {Peretto}, N. and {Men'shchikov}, A. and {Ward-Thompson}, D. and {Kirk}, J. and {Griffin}, M. and {Marsh}, K. and {Abergel}, A. and {Arzoumanian}, D. and {Benedettini}, M. and {Hill}, T. and {Motte}, F. and {Nguyen Luong}, Q. and {Pezzuto}, S. and {Rivera-Ingraham}, A. and {Roussel}, H. and {Rygl}, K.~L.~J. and {Spinoglio}, L. and {Stamatellos}, D. and {White}, G.},
        title = "{Reconstructing the density and temperature structure of prestellar cores from Herschel data: A case study for B68 and L1689B}",
      journal = {\aap},
         year = 2014,
        month = feb,
       volume = {562},
          eid = {A138},
        pages = {A138},
          doi = {10.1051/0004-6361/201322236},
archivePrefix = {arXiv},
       eprint = {1311.5086},
 primaryClass = {astro-ph.GA},
       adsurl = {https://ui.adsabs.harvard.edu/abs/2014A&A...562A.138R}
}

@ARTICLE{Scibelli2023,
       author = {{Scibelli}, Samantha and {Shirley}, Yancy and {Schmiedeke}, Anika and {Svoboda}, Brian and {Singh}, Ayushi and {Lilly}, James and {Caselli}, Paola},
        title = "{3D radiative transfer modelling and virial analysis of starless cores in the B10 region of the Taurus molecular cloud}",
      journal = {\mnras},
         year = 2023,
        month = may,
       volume = {521},
       number = {3},
        pages = {4579-4597},
          doi = {10.1093/mnras/stad827},
archivePrefix = {arXiv},
       eprint = {2303.09574},
 primaryClass = {astro-ph.GA},
       adsurl = {https://ui.adsabs.harvard.edu/abs/2023MNRAS.521.4579S}
}

@ARTICLE{Palau2021,
       author = {{Palau}, Aina and {Zhang}, Qizhou and {Girart}, Josep M. and {Liu}, Junhao and {Rao}, Ramprasad and {Koch}, Patrick M. and {Estalella}, Robert and {Chen}, Huei-Ru Vivien and {Liu}, Hauyu Baobab and {Qiu}, Keping and {Li}, Zhi-Yun and {Zapata}, Luis A. and {Bontemps}, Sylvain and {Ho}, Paul T.~P. and {Beuther}, Henrik and {Ching}, Tao-Chung and {Shinnaga}, Hiroko and {Ahmadi}, Aida},
        title = "{Does the Magnetic Field Suppress Fragmentation in Massive Dense Cores?}",
      journal = {\apj},
         year = 2021,
        month = may,
       volume = {912},
       number = {2},
          eid = {159},
        pages = {159},
          doi = {10.3847/1538-4357/abee1e},
archivePrefix = {arXiv},
       eprint = {2010.12099},
 primaryClass = {astro-ph.GA},
       adsurl = {https://ui.adsabs.harvard.edu/abs/2021ApJ...912..159P}
}

@ARTICLE{Li2007,
       author = {{Li}, D. and {Velusamy}, T. and {Goldsmith}, P.~F. and {Langer}, William D.},
        title = "{Massive Quiescent Cores in Orion. II. Core Mass Function}",
      journal = {\apj},
         year = 2007,
        month = jan,
       volume = {655},
       number = {1},
        pages = {351-363},
          doi = {10.1086/509736},
archivePrefix = {arXiv},
       eprint = {astro-ph/0610634},
 primaryClass = {astro-ph},
       adsurl = {https://ui.adsabs.harvard.edu/abs/2007ApJ...655..351L}
}

@ARTICLE{Tatematsu2010,
       author = {{Tatematsu}, Ken'ichi and {Hirota}, Tomoya and {Kandori}, Ryo and {Umemoto}, Tomofumi},
        title = "{Chemical Variation in Molecular Cloud Cores in the Orion A Cloud}",
      journal = {\pasj},
         year = 2010,
        month = dec,
       volume = {62},
        pages = {1473},
          doi = {10.1093/pasj/62.6.1473},
archivePrefix = {arXiv},
       eprint = {1010.4939},
 primaryClass = {astro-ph.GA},
       adsurl = {https://ui.adsabs.harvard.edu/abs/2010PASJ...62.1473T}
}

@ARTICLE{Stanke2002,
       author = {{Stanke}, T. and {McCaughrean}, M.~J. and {Zinnecker}, H.},
        title = "{An unbiased H$_{2}$ survey for protostellar jets in Orion A. II\textbackslash@. The infrared survey data}",
      journal = {\aap},
         year = 2002,
        month = sep,
       volume = {392},
        pages = {239-266},
          doi = {10.1051/0004-6361:20020763},
       adsurl = {https://ui.adsabs.harvard.edu/abs/2002A&A...392..239S}
}

@ARTICLE{Tatematsu2008,
       author = {{Tatematsu}, Ken'ichi and {Kandori}, Ryo and {Umemoto}, Tomofumi and {Sekimoto}, Yutaro},
        title = "{N$_{2}$H$^{+}$ and HC$_{3}$N Observations of the Orion A Cloud}",
      journal = {\pasj},
         year = 2008,
        month = jun,
       volume = {60},
        pages = {407},
          doi = {10.1093/pasj/60.3.407},
archivePrefix = {arXiv},
       eprint = {0804.0111},
 primaryClass = {astro-ph},
       adsurl = {https://ui.adsabs.harvard.edu/abs/2008PASJ...60..407T}
}

@ARTICLE{Schnee2005,
       author = {{Schnee}, S. and {Goodman}, A.},
        title = "{Density and Temperature Structure of TMC-1C from 450 and 850 Micron Maps}",
      journal = {\apj},
         year = 2005,
        month = may,
       volume = {624},
       number = {1},
        pages = {254-266},
          doi = {10.1086/429156},
archivePrefix = {arXiv},
       eprint = {astro-ph/0502024},
 primaryClass = {astro-ph},
       adsurl = {https://ui.adsabs.harvard.edu/abs/2005ApJ...624..254S}
}

@software{Dullemond2012,
       author = {{Dullemond}, C.~P. and {Juhasz}, A. and {Pohl}, A. and {Sereshti}, F. and {Shetty}, R. and {Peters}, T. and {Commercon}, B. and {Flock}, M.},
        title = "{RADMC-3D: A multi-purpose radiative transfer tool}",
 howpublished = {Astrophysics Source Code Library, record ascl:1202.015},
         year = 2012,
        month = feb,
          eid = {ascl:1202.015},
archivePrefix = {ascl},
       eprint = {1202.015},
       adsurl = {https://ui.adsabs.harvard.edu/abs/2012ascl.soft02015D}
}

@ARTICLE{Miettinen2020,
       author = {{Miettinen}, O.},
        title = "{Dense cores in the Seahorse infrared dark cloud: physical properties from modified blackbody fits to the far-infrared-submillimetre spectral energy distributions}",
      journal = {\aap},
         year = 2020,
        month = dec,
       volume = {644},
          eid = {A82},
        pages = {A82},
          doi = {10.1051/0004-6361/202039204},
archivePrefix = {arXiv},
       eprint = {2011.04293},
 primaryClass = {astro-ph.GA},
       adsurl = {https://ui.adsabs.harvard.edu/abs/2020A&A...644A..82M}
}

@ARTICLE{Stamatellos2010,
       author = {{Stamatellos}, D. and {Griffin}, M.~J. and {Kirk}, J.~M. and {Molinari}, S. and {Sibthorpe}, B. and {Ward-Thompson}, D. and {Whitworth}, A.~P. and {Wilcock}, L.~A.},
        title = "{Modelling Herschel observations of infrared-dark clouds in the Hi-GAL survey}",
      journal = {\mnras},
         year = 2010,
        month = nov,
       volume = {409},
       number = {1},
        pages = {12-21},
          doi = {10.1111/j.1365-2966.2010.17093.x},
archivePrefix = {arXiv},
       eprint = {1006.1390},
 primaryClass = {astro-ph.GA},
       adsurl = {https://ui.adsabs.harvard.edu/abs/2010MNRAS.409...12S}
}

@ARTICLE{Steinacker2005,
       author = {{Steinacker}, J. and {Bacmann}, A. and {Henning}, Th. and {Klessen}, R. and {Stickel}, M.},
        title = "{3D continuum radiative transfer in complex dust configurations.  II. 3D structure of the dense molecular cloud core {\ensuremath{\rho}} Oph D}",
      journal = {\aap},
         year = 2005,
        month = apr,
       volume = {434},
       number = {1},
        pages = {167-180},
          doi = {10.1051/0004-6361:20041978},
archivePrefix = {arXiv},
       eprint = {astro-ph/0410635},
 primaryClass = {astro-ph},
       adsurl = {https://ui.adsabs.harvard.edu/abs/2005A&A...434..167S}
}

@ARTICLE{Schnee2007_0,
       author = {{Schnee}, S. and {Kauffmann}, J. and {Goodman}, A. and {Bertoldi}, F.},
        title = "{The Effect of Noise in Dust Emission Maps on the Derivation of Column Density, Temperature, and Emissivity Spectral Index}",
      journal = {\apj},
         year = 2007,
        month = mar,
       volume = {657},
       number = {2},
        pages = {838-848},
          doi = {10.1086/511054},
archivePrefix = {arXiv},
       eprint = {astro-ph/0611535},
 primaryClass = {astro-ph},
       adsurl = {https://ui.adsabs.harvard.edu/abs/2007ApJ...657..838S}
}

@ARTICLE{Schnee2007,
       author = {{Schnee}, S. and {Caselli}, P. and {Goodman}, A. and {Arce}, H.~G. and {Ballesteros-Paredes}, J. and {Kuchibhotla}, K.},
        title = "{TMC-1C: An Accreting Starless Core}",
      journal = {\apj},
         year = 2007,
        month = dec,
       volume = {671},
       number = {2},
        pages = {1839-1857},
          doi = {10.1086/521577},
archivePrefix = {arXiv},
       eprint = {0706.4115},
 primaryClass = {astro-ph},
       adsurl = {https://ui.adsabs.harvard.edu/abs/2007ApJ...671.1839S}
}

@ARTICLE{Schnee2010,
       author = {{Schnee}, Scott and {Enoch}, Melissa and {Noriega-Crespo}, Alberto and {Sayers}, Jack and {Terebey}, Susan and {Caselli}, Paola and {Foster}, Jonathan and {Goodman}, Alyssa and {Kauffmann}, Jens and {Padgett}, Deborah and {Rebull}, Luisa and {Sargent}, Anneila and {Shetty}, Rahul},
        title = "{The Dust Emissivity Spectral Index in the Starless Core TMC-1C}",
      journal = {\apj},
         year = 2010,
        month = jan,
       volume = {708},
       number = {1},
        pages = {127-136},
          doi = {10.1088/0004-637X/708/1/127},
archivePrefix = {arXiv},
       eprint = {0911.0892},
 primaryClass = {astro-ph.GA},
       adsurl = {https://ui.adsabs.harvard.edu/abs/2010ApJ...708..127S}
}

@ARTICLE{Koley2022,
       author = {{Koley}, Atanu},
        title = "{Studying the chemical and kinematical structures of dense cores TMC-1C, L1544, and TMC-1 in the Taurus molecular cloud using CCS and NH$_{3}$ observations}",
      journal = {\mnras},
         year = 2022,
        month = oct,
       volume = {516},
       number = {1},
        pages = {185-196},
          doi = {10.1093/mnras/stac1935},
archivePrefix = {arXiv},
       eprint = {2208.00968},
 primaryClass = {astro-ph.GA},
       adsurl = {https://ui.adsabs.harvard.edu/abs/2022MNRAS.516..185K}
}

@ARTICLE{Crapsi2005,
       author = {{Crapsi}, A. and {Caselli}, P. and {Walmsley}, C.~M. and {Myers}, P.~C. and {Tafalla}, M. and {Lee}, C.~W. and {Bourke}, T.~L.},
        title = "{Probing the Evolutionary Status of Starless Cores through N$_{2}$H$^{+}$ and N$_{2}$D$^{+}$ Observations}",
      journal = {\apj},
         year = 2005,
        month = jan,
       volume = {619},
       number = {1},
        pages = {379-406},
          doi = {10.1086/426472},
archivePrefix = {arXiv},
       eprint = {astro-ph/0409529},
 primaryClass = {astro-ph},
       adsurl = {https://ui.adsabs.harvard.edu/abs/2005ApJ...619..379C}
}


\bsp
\label{lastpage}
\end{document}